\documentclass[journal,twoside]{IEEEtran}
\usepackage{mathrsfs}
\usepackage{amsbsy}
\usepackage{graphicx}
\usepackage{xcolor}
\usepackage{amsmath}
\usepackage{amsthm}
\usepackage{amssymb}
\allowdisplaybreaks

\usepackage{nomencl}
\usepackage{commath}
\usepackage[T1]{fontenc}
\usepackage{mathtools}
\usepackage{float}
\usepackage{url}
\usepackage{algorithm}
\usepackage{pifont}
\usepackage{algpseudocode}
\usepackage{etoolbox}
\usepackage{lipsum}

\usepackage{subfigure}
\usepackage{hhline}
\usepackage{multirow}

\usepackage[shortlabels]{enumitem}
\usepackage{blindtext}
\usepackage{wasysym}
\usepackage{balance}
\usepackage{array}

\usepackage{makecell}

\makenomenclature

\usepackage{cite}

\ifCLASSINFOpdf
\else
\fi

\begin{document}
%
\title{Chance-Constrained Economic Dispatch with Flexible Loads and RES}
%
%
%


\author{Tian~Liu,~\IEEEmembership{Student Member,~IEEE,}\thanks{T. Liu and D.H.K. Tsang are with the Hong Kong University of Science and Technology, Hong Kong (e-mail: \{tliuai, eetsang\}@ust.hk). 

B. Sun is with School of Computer Science, University of Waterloo, Canada (e-mail: bo.sun@@uwaterloo.ca)} Bo~Sun,~\IEEEmembership{Member,~IEEE,}  and~Danny~H.K.~Tsang,~\IEEEmembership{Fellow,~IEEE}}


%
%


\markboth{}{}

%



\maketitle

\begin{abstract}
 With the increasing penetration of intermittent renewable energy sources (RESs), it becomes increasingly challenging to maintain the supply-demand balance of power systems by solely relying on the generation side. To combat the volatility led by the uncertain RESs, demand-side management by leveraging the multi-dimensional flexibility (MDF) has been recognized as an economic and efficient approach. Thus, it is important to integrate MDF into existing power systems. In this paper, we propose an enhanced day-ahead energy market, where the MDFs of aggregate loads are traded to minimize the generation cost and mitigate the volatility of locational marginal prices (LMPs) in the transmission network. We first explicitly capture the negative impact of the uncertainty from RESs on the day-ahead market by a chance-constrained economic dispatch problem (CEDP). Then, we propose a bidding mechanism for the MDF of the aggregate loads and combine this mechanism into the CEDP for the day-ahead market. Through multiple case studies, we show that MDF from load aggregators can reduce the volatility of LMPs. In addition, we identify the values of the different flexibilities in the MDF bids, which provide useful insights into the design of more complex MDF markets.

\end{abstract}


%
\IEEEpeerreviewmaketitle

\section{Introduction}
%
%
%
%

\IEEEPARstart{A}{s} more and more renewable energy sources (RESs) are integrated into the grid, it becomes increasingly challenging and costly to maintain the balance between generation and consumption by purely relying on the supply side due to the increasing volatility resulting from intermittent generation of RESs. In this regard, demand-side management (DSM) by leveraging the flexibilities of the electricity loads is one promising scheme to overcome the limits of traditional power systems, in which the flexibility of the loads is, however, not fully exploited to improve the grid's operational efficiency.

A majority of literature on DSM have put their efforts in utilizing the flexibilities of controllable loads to facilitate the operation of the power system. In these works, various types of load flexibilities have been considered for different purposes. For example, authors in \cite{ramirez2016co,gottwalt2016modeling} focus on reducing the generation cost by the shifting flexibility, which allows a fixed amount of load to be shifted over a certain time period. In \cite{ramirez2016co}, a planning model that allows co-optimizing the capital and operational cost of conventional generation assets as well as the shifting flexibility of electric vehicles (EVs) is proposed. The authors demonstrate that the flexibility from EVs can effectively flatten net-load, namely the load minus the wind and solar generation, by reducing peak demand levels. In addition, the value of the EV flexibility in terms of the total system cost saving has been shown to increase with an increasing electrification of the transport sector and increasing wind generation capacity levels, which highlights the potential of flexible loads for the system cost reduction. In \cite{gottwalt2016modeling}, a centralized scheduling model to exploit demand flexibilities from residential devices is provided to reduce generation costs of a microgrid with a large share of renewable generation. It is shown that the generation cost reduction can be achieved by contracting a large number of flexible loads such as EVs and storage water heaters.
Moreover, authors in \cite{negrete2016rate} discuss power-rating flexibility which is characterized by the total amount of energy that must be delivered over a specified time period and the maximum rate at which this energy may be delivered. The authors also address the problem of finding optimal market decisions for a supplier that can provide such flexibility in a forward market. In addition, Bitar \textit{et al.} \cite{bitar2017deadline} studies the deadline flexibility, where consumers consent to
deferred service of prespecified loads in exchange for a reduced
price for energy. The authors design a pricing scheme that yields
an efficient competitive equilibrium between the suppliers and
consumers. Furthermore, Nayyar \textit{et al.} \cite{nayyar2016duration} considers the duration flexibility, in which power may be delivered at any time so long as the total duration of service is equal to
the load's specified duration.

We refer to all kinds of flexibilities mentioned above as the multi-dimensional flexibility (MDF) in the rest of the paper. Although the contributions of a single type of load flexibility has been well-studied in the literature, there is a lack of a unified framework to combine MDFs to the power system operation. Moreover, to encourage the participation of controllable loads into the power system operation, the system operator needs to design the underlying electricity market to provide sufficient incentives for the flexible loads.

In the literature, various new markets have been proposed for trading the load flexibilities between the load aggregator and small-size flexible loads \cite{negrete2016rate,bitar2017deadline,nayyar2016duration}. In these markets, small-size loads are typically given a reduced per unit energy price by offering their power-rating flexibility \cite{negrete2016rate}, deadline flexibility \cite{bitar2017deadline}, and duration flexibility \cite{nayyar2016duration}. 
The aggregator trades the aggregate flexibilities in the transmission-level electricity market to gain net benefit for the designed market, and shares the benefit with the flexible loads by designing the reduced energy price.



Complementary to existing works, this paper focuses on the power system operation and the transmission-level market design for the MDFs of load aggregators. To be specific, this work can be considered as an enhancement of the day-ahead market when MDFs from controllable loads are combined into the market clearing process. Particularly, the MDF service from each aggregator is composed of the following key attributes: i) a specified service period (e.g., 10 a.m. to 4 p.m.); ii) the maximal load adjustment with respect to the base load reference (normal consumption) during the service period and iii) the maximal load adjustment per unit time slot (e.g., an hour). Moreover, since the transmission network constraints are taken into account, the locational information of the service is also captured. 

Given that this paper focuses on the power system operation and market design of the transmission network level, we consider the MDF from the aggregate loads on each bus. 
To characterize the features of MDF, we model the aggregate load of each bus as a virtual battery (VB) following the previous works \cite{hao2015generalized,hao2015aggregate,hughes2016identification,zhao2017geometric,8081791}, and all the MDFs are quantified by the parameters of the VBs. The VB model is a scalar linear system that resembles a simplified battery dynamics parameterized by charge and discharge power limits, energy capacity limits, and self-discharge rate. Details about estimating the parameters of the VBs can be found in \cite{hao2015generalized,hughes2016identification,zhao2017geometric}.

The challenges of integrating the MDFs into the current day-ahead market mainly lie in two aspects. Firstly, a bidding process for the load aggregators needs to be designed so that the heterogeneous MDFs from different aggregators can be utilized in the market clearing process and the profit of each aggregator can be determined accordingly. 
Secondly, the uncertain output of RESs, as the key driving factor of the integration of MDFs, needs to be taken into account in the design of the enhanced day-ahead market.
In the following, we will start dealing with the second challenge by explicitly modeling the uncertain output of RESs and chance constraints are added into the economic dispatch (ED) problem in the traditional market clearing process. 
Then, based on the chance-constrained economic dispatch problem (CEDP), the bidding scheme for the MDFs is designed and an efficient algorithm for the whole market clearing problem is proposed.
In summary, the main contributions of this work are as follows:


\begin{enumerate}
	\item We propose a framework to integrate the load aggregators with MDFs into the day-ahead market. A unified representation of heterogeneous MDFs based on a VB model is adopted and a bidding scheme is designed to integrate MDFs into the CEDP by parametric reward bidding. 
	\item Our proposed market clearing problem is successfully transformed into an SOCP, which facilitates the computation of solving the optimization problem.
	\item The reliability of our models is justified by simulating different distributions of wind turbines' output.
\end{enumerate}


\section{Chance-constrained ED Problem}\label{mdf:sec2}
In this section, we introduce the chance-constrained economic dispatch problem (CEDP), which enhances the current day-ahead market by modeling the uncertainties from RESs into a set of chance constraints. Based on the CEDP, we can evaluate the impact of uncertain RESs on the day-ahead market effectively.

\subsection{Wind Uncertainty Model and Affine Control Policy}
We consider a transmission network with $N$ buses and denote the set of all the buses by $\mathcal{B}:=\{1,2,\cdots,N\}$. Let $\mathcal{W} \subseteq \mathcal{B}$ be the set of buses connected to wind turbines. The uncertain wind power output of all the buses in $\mathcal{W}$ at time $t$ is denoted by the $n_{\omega} \times 1$ vector $\boldsymbol{P}_{\omega,t}$, where $n_{\omega}$ is the total number of buses equipped with wind turbines. Furthermore, we model $\boldsymbol{P}_{\omega,t}$ in the mean-plus-noise form as follows,
\begin{equation}
\boldsymbol{P}_{\omega,t}=\overline{\boldsymbol{P}}_{\omega,t}+\boldsymbol{\omega}_t,
\end{equation}
where $\overline{\boldsymbol{P}}_{\omega,t}$ is the known expected wind outputs\footnote{Note that $\overline{\boldsymbol{P}}_{\omega,t}$ can be obtained by the state-of-the-art forecast methods but the details of derivation is out of the scope of this paper.} and $\boldsymbol{\omega}_t$ denotes the uncertain deviations from the expected values. Moreover, $\boldsymbol{\omega}_t$ is assumed to follow the normal distribution with zero mean and known covariance matrix {\color{black}$\Sigma_t \in \mathbb{R}^{n_{\omega}\times n_{\omega}}$, where $\mathbb{R}$ denotes the set of real numbers}. Let $\boldsymbol{1}_{x}$ denote the $x \times 1$ vector with all elements equal to one. Then the total deviation of the wind power output is $\Omega_t=\boldsymbol{1}_{n_{\omega}}^T\boldsymbol{\omega}_t$. Because $\Omega_t$ is a linear combination of Gaussian variables, $\Omega_t$ follows normal distribution with zero mean and variance $\sigma_t^2=\boldsymbol{1}_{n_{\omega}}^T\Sigma_t\boldsymbol{1}_{n_{\omega}}$.

\begin{figure}[t!]
	\centering     
	\subfigure[At 19:00 of each day.]{\label{mdfcaiso_fig:02a1}\includegraphics[width=42mm]{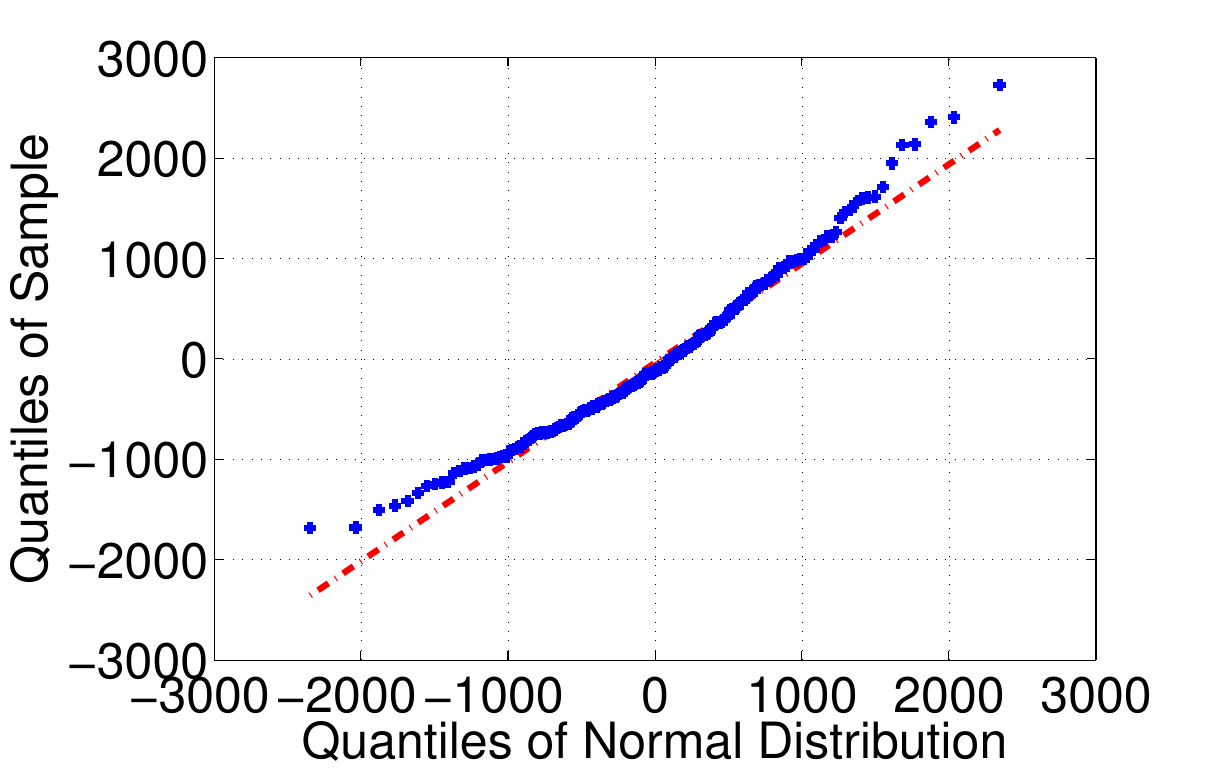}}
	\subfigure[At 20:00 of each day.]{\label{mdfcaios_fig:02a2}\includegraphics[width=42mm]{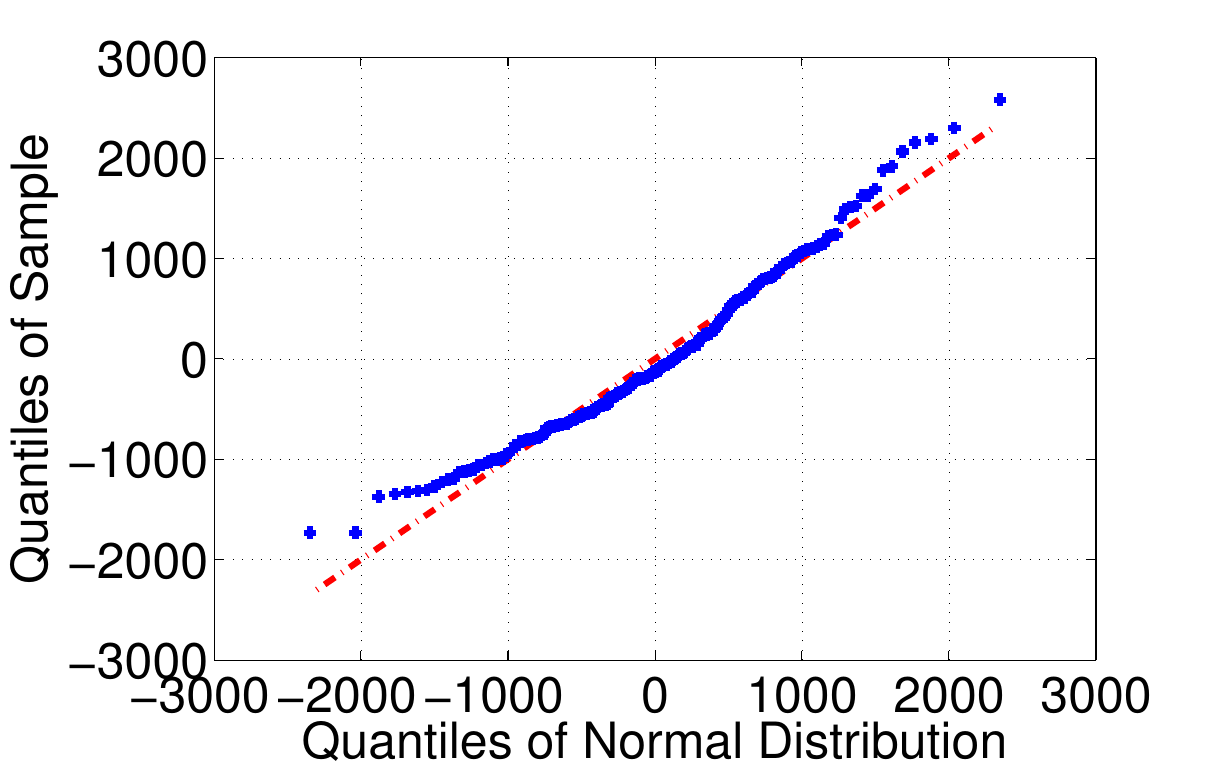}}
	\caption{Q-Q plots for hourly forecast errors of wind power output in August from 2011 to 2017.}\label{mdfcaios_fig:02a}
\end{figure}

Note that the assumption of normality has been widely used for modeling the uncertain wind turbine output deviation \cite{bienstock2014chance,roald2017corrective,lubin2016robust}. We justify this by analyzing some real data of actual renewable production from California Independent System Operator (CAISO) \cite{websitecaiso}. Specifically, we use quantile-quantile plots (Q-Q plots) which plot the quantiles (values that split a data set into equal portions) of the data set against that of the corresponding Gaussian distribution to check the normality visually \cite{ghasemi2012normality}. In Fig. \ref{mdfcaios_fig:02a}, we show the Q-Q plots for hourly forecast errors of wind power outputs at 19:00 and 20:00 in August across seven years (from 2011 to 2017). From the figure, we can see that the data generally follow the normal distribution except at the two ends. However, this normality assumption is imposed for mathematical simplicity, which can be relaxed or refined easily. Additionally, for cases where the assumption of a normal distribution is not justified, our proposed framework can be easily extended to the cases where only limited knowledge of the distribution is available \cite{zhang2017distributionally,summers2014stochastic,roald2015security,xie2017distributionally}.

Let $\mathcal{G}$ denote the set of all the traditional generators in the system, and $\boldsymbol{P}_{g,t} := \{P_{g,i,t}, i\in \cal G \}$ denote the vector of power outputs from these generators. Since the power outputs of wind turbines are uncertain, the outputs of the traditional generators need to be adjusted in real-time to ensure the matching of supply and demand. To model the interaction between the outputs of the wind power and traditional power, we assume that all traditional generators follow an affine control policy in real-time to compensate for the deviation of wind power outputs. Specifically, the affine control policy means that at time slot $t$, the total wind power deviation is allocated to the controllable generators in a proportional way, and thus the actual power output  vector of the dispatchable generators can be described by:
\begin{equation}
\boldsymbol{P}_{g,t}=\overline{\boldsymbol{P}}_{g,t}-\boldsymbol{\beta}_{g,t}\cdot \Omega_t,\label{eq:affine_g1}
\end{equation} 
where $\overline{\boldsymbol{P}}_{g,t}:= \{\overline{P}_{g,i,t}, \forall i \in \cal G \}$ denotes the vector of the set-points for the generator outputs and $\boldsymbol{\beta}_{g,t}$ is the vector of the allocation coefficients. In order to maintain the balance between the supply and demand, the following constraints are imposed:
\begin{equation}
\begin{aligned}
\boldsymbol{\beta}_{g,t} \geq 0, \ \boldsymbol{1}_{n_g}^T\boldsymbol{\beta}_{g,t}=1,\label{mdf_eq_aff_g}
\end{aligned}
\end{equation}
where $n_g$ is the total number of dispatchable generators. Note that $\boldsymbol{P}_{g,t}$ is also random due to its relation with $\Omega_t$ in Eq. (\ref{eq:affine_g1}).

\subsection{CEDP for Incorporating Uncertain Generation}
The randomness of the wind generators enforces the day-ahead market to solve a CEDP, which differs from traditional ED problems from two aspects. Firstly, the objective of the CEDP is to minimize the \textit{expected} generation cost. Secondly, the operational constraints of the traditional ED problems can not be imposed in a deterministic way. By contrast, chance constraints are employed in the CEDP such that the probability of violating the operational constraints is guaranteed to be lower than a small value {\color{black}$\epsilon$ and we refer to it as a risk parameter}. Specifically, the output of generator $i$, should be within the corresponding limits $P_{g,i}^{max}$ and $P_{g,i}^{min}$:
\begin{align}
&	\mathbb{P}_{\Omega_t}[P_{g,i,t}\leq P_{g,i}^{max}]\geq 1-\epsilon^{g}_{i},\forall i \in \mathcal{G},\label{mdf:ed_cst_g1} \\
&	\mathbb{P}_{\Omega_t}[P_{g,i,t}\geq P_{g,i}^{min}]\geq 1-\epsilon^{g}_{i},\forall i \in \mathcal{G}, \label{mdf:ed_cst_g2}
\end{align}
In addition, the power flow on each transmission line should be below the maximum transmission capacity. Let $\cal L$ denote the set of all the transmission lines {\color{black}with a cardinality of $L$}. According to the DC power flow equations, the power flow on transmission line $l$ at time $t$, which is denoted by $F_{l,t}$, is a linear combination of the generation of all the generators and consumptions of all the loads. Let $\cal D \subseteq \cal B$ be the set of load buses, and $\boldsymbol{P}_{d,t} := \{P_{d,i,t}, i \in \cal D \}$ denote the vector of base loads. Then, the vector of the power flows on all the transmission lines  $\boldsymbol{F}_t := \{F_{l,t}, l \in \cal L \}$ is determined by
\begin{equation}
\boldsymbol{F}_t=\boldsymbol{\Gamma}(H_g\boldsymbol{P}_{g,t}+H_w\boldsymbol{P}_{w,t}-H_d\boldsymbol{P}_{d,t}),\label{mdf:ed_lineflow_1}
\end{equation}
where $H_g$, $H_{\omega}$ and $H_d$ denote the bus connection matrix for dispatchable generator buses, wind turbine buses and non-flexible load buses, respectively. For example, $H_g\in R^{N\times n_g}$ is constructed such that its $(i,j)$-th element is 1 if and only if generator $j$ is located at bus $i$, while all the other elements are zero. In addition, {\color{black}$\boldsymbol{\Gamma} \in \mathbb{R}^{L \times N}$} denotes the matrix of generation shift factors. To avoid the overloading of the transmission lines, we impose the following chance constraints
\begin{align}
&  \mathbb{P}_{\boldsymbol{\omega}_t}[	F_{l,t}\leq F_{l}^{max}]\geq 1-\epsilon_{l},\forall l \in \mathcal{L}, \label{mdf:ed_cst_l1}\\
&  \mathbb{P}_{\boldsymbol{\omega}_t}[	F_{l,t}\geq -F_{l}^{max}]\geq 1-\epsilon_{l},\forall l \in \mathcal{L}.\label{mdf:ed_cst_l2}
\end{align}

Then, for each time slot $t$, the CEDP is formulated as follows:

\begin{align}
& \underset{\overline{\boldsymbol{P}}_{g,t},{\boldsymbol{\beta}}_{g,t}}{\text{min}}
& & \mathbb{E}\Big(\sum_{i\in\mathcal{G}}C_i(P_{g,i,t})\Big)\\
& \text{s. t.}
& & \boldsymbol{1}_{n_g}^T\overline{\boldsymbol{P}}_{g,t}+\boldsymbol{1}_{n_w}^T\overline{\boldsymbol{P}}_{w,t}=\boldsymbol{1}_{n_d}^T\boldsymbol{P}_{d,t},  \label{mdf:ed_cst_1} \\
& & &  \textrm{constraints} \ (\ref{eq:affine_g1})-(\ref{mdf:ed_cst_l2}).\nonumber
\end{align}

In this problem, constraint (\ref{mdf:ed_cst_1}) requires that the total demand should be balanced by the summation of the dispatchable generator outputs and the wind generation in the sense of expectation, where $n_d$ denotes the number of load buses. Meanwhile, under the uncertainty of wind power, it is guaranteed that the supply can exactly match the demand through the affine control policy defined by Eq. (\ref{eq:affine_g1}). 

This CEDP can be used to evaluate the impact of the wind generation on a traditional ED problem.  Furthermore, by incorporating the controllable loads into the CEDP in the affine control policy, the generation cost is expected to decrease. Moreover, a market is needed to ensure the participation of the controllable loads. In this regard, we will introduce the market clearing problem when controllable loads with MDF are integrated into the CEDP.

\section{Market Clearing Problem for MDF Services}\label{mdf:sec3}

We start this section by introducing several key features of our market clearing framework for MDF services. First, we assume that at load buses in a transmission network, there may exist MDF aggregators which represent the aggregate flexibilities of load buses by the VB model \cite{hao2015generalized,hao2015aggregate,hughes2016identification,zhao2017geometric,8081791}. Basically, the flexible load at each bus can be modeled by a battery which is characterized by an energy capacity and the charge/discharge rate limits. The detailed {\color{black}demand response (DR)} aggregation method can be found in \cite{hao2015generalized,hao2015aggregate,hughes2016identification,zhao2017geometric,8081791}. 
	Second, for each aggregator, apart from submitting the key parameters of its VB model, it should also submit a \textit{parametric reward function} which is used for characterizing the remuneration requested by the aggregator. The form of this parametric reward function is defined by the market operator and the aggregator only needs to submit the parameters of its reward function. Different from a simple bid consisting of a price-volume pair in a traditional market, in our framework, the reward is a function of the actual flexibilities accepted by the market operator. Then, with the MDF bids from aggregators, the market clearing process is integrated into the traditional ED problem and the objective is to minimize the summation of the total expected generation cost and the total payment for the MDF services of the load aggregators. One important feature of our clearing process is that the flexibility bids can be \textit{partially} accepted. This means that a portion of the flexibility service can be accepted. In contrast to a traditional DR market where a bid can either be totally accepted or rejected, our market design will bring more flexibilities to both the operator and aggregators. In addition, the actual payment is dependent on the actual service that is accepted. Therefore, the evaluation for the complex MDF services is solved by this bidding-clearing process. Moreover, we reformulate the chance-constrained market clearing problem into an SOCP such that it can be solved efficiently.

\subsection{MDF Bidding for Incorporating Controllable Loads}
Consider the market clearing problem for a time horizon of $\mathcal{T}:=\{1,2,\cdots,T\}$. Denote the set of buses that will offer the MDF services as $\mathcal{F} \subseteq \mathcal{B}$. Assume that on each bus $i \in \cal F$, there exists one load aggregator, and its MDF bid is described as
$
	\boldsymbol{\psi}_i=\big\{t_i^s,t_i^e,R_i^{min},R_i^{max},E_i^{min},E^{max}_i;\gamma_i^R,\gamma_i^E\big\}.
$
There are two parts in this MDF bid. The first part is related to the parameters of the VB model, which abstracts the MDFs of the aggregate loads. Specifically,  $t_i^s \in \mathcal{T}$ and $t_i^e\in\mathcal{T}$ define the available service period $[t_i^s,t_i^e]$ of the controllable load. $R_i^{min} \leq 0$ and $R_i^{max} \geq 0$ denote the maximal charging and discharging rates of the VB. Additionally, $E_i^{min}\leq 0$ and $E_i^{max}\geq 0$ represent the maximal energy that can be released from or stored into the VB, respectively. Note that, releasing and storing energy of the VB model are with respect to the base load. For example, for the load bus $i$ with controllable loads, if an amount of $\Delta D>0$ energy is released from the aggregate load, the actual load consumption is $P_{d,i,t}-\Delta D$, which means the load is decreased by $\Delta D$. More details about abstracting aggregate flexibilities as VB can be found in  \cite{hao2015generalized,hao2015aggregate,hughes2016identification,zhao2017geometric,8081791}.

Recall that the market operator is allowed to accept only a portion of each MDF bid. Therefore, four decision variables are associated with each $\boldsymbol{\psi}_i$ in the market clearing problem and they are denoted by: $
\boldsymbol{\alpha}_i=(\alpha_i^{R-},\alpha_i^{R+},\alpha_i^{E-},\alpha_i^{E+})
$. $\boldsymbol{\alpha}_i$ satisfies the following constraints:
\begin{align}
& R_i^{min} \leq \alpha_i^{R-} \leq 0 \leq \alpha_i^{R+} \leq R_i^{max},\label{mdf:alpha_1}\\
& E_i^{min} \leq \alpha_i^{E-} \leq 0 \leq \alpha_i^{E+} \leq E_i^{max}.\label{mdf:alpha_2}
\end{align}
Different from the VB parameters in the MDF bid, $\gamma_i^P$ and $\gamma_i^E$ are the valuation factors that are determined by each load aggregator based on their own evaluation of their MDF. Based on these valuation factors and the market clearing results, the market operator pays each load aggregator a reward $r_i(\boldsymbol{\alpha_i})$. A simple example of the parametric reward can be a linear function:
$
r_i(\boldsymbol{\alpha}_i)=\gamma_i^P(\alpha_i^{R+}-\alpha_i^{R-})+\gamma_i^E(\alpha_i^{E+}-\alpha_i^{E-}),
$
where $\gamma_i^P$ and $\gamma_i^E$ are the linear coefficients for the total amount of the flexible power and energy, respectively. The form of this parametric reward function is defined by the operator to facilitate the clearing process of the MDF market and it is not limited to a linear function. Our following results can be extended easily for any convex function of $r_i(\boldsymbol{\alpha}_i)$. 

With the participation of MDF services, the output deviations from the wind turbines can be compensated by both the dispatchable generators and load aggregators. Thus, similar to Eq. (\ref{eq:affine_g1}), the affine control policy can also be applied to the discharging power of the equivalent VBs from the load aggregators:
\begin{equation}
\boldsymbol{P}_{f,t}=\overline{\boldsymbol{P}}_{f,t}-\boldsymbol{\beta}_{f,t}\cdot \Omega_t,\label{eq:affine_f}
\end{equation}
where $\overline{\boldsymbol{P}}_{f,t}$ denotes the vector of the set-points for the discharging power of VBs\footnote{{\color{black}When $\overline{\boldsymbol{P}}_{f,t}\leq 0$, it refers to the charging of the VB at the set-point.}} and $\boldsymbol{\beta}_{f,t}$ denotes the corresponding allocation coefficients. Accordingly, the allocation coefficients $\boldsymbol{\beta}_{g,t}$ and $\boldsymbol{\beta}_{f,t}$ should be constrained as:
\begin{equation}
\boldsymbol{\beta}_{g,t} \geq 0, \ \boldsymbol{\beta}_{f,t} \geq 0,\ \boldsymbol{1}_{n_g}^T\boldsymbol{\beta}_{g,t}+\boldsymbol{1}_{n_f}^T\boldsymbol{\beta}_{f,t}=1,\forall t \in \cal T,\label{mdf:for beta1}
\end{equation}
where $n_f$ denotes the number of load aggregators. Also, constraints for the evolution of state-of-charge (SOC) of VBs are imposed as:
\begin{equation}
	\boldsymbol{E}_0=\boldsymbol{0},\ 	\boldsymbol{E}_{t+1}=	\boldsymbol{E}_{t}-\boldsymbol{P}_{f,t},\forall t \in \cal T, \label{mdf:cst_e2}
\end{equation}
where $\boldsymbol{E}_{t}:=\{E_{i,t},i\in \mathcal{F}\}$ denotes the vector of the SOC of VBs.
Furthermore, the power flow Eq. (\ref{mdf:ed_lineflow_1}) is adjusted as:
\begin{equation}
 \boldsymbol{F}_t=\boldsymbol{\Gamma}(H_g\boldsymbol{P}_{g,t}+H_w\boldsymbol{P}_{w,t}+H_f\boldsymbol{P}_{f,t}-H_d\boldsymbol{P}_{d,t}),\label{mdf:lineflow_1}
 \end{equation} 
where $H_f$ denotes the bus connection matrix for load aggregator buses with MDF bids. 

\subsection{Market Clearing Problem by the CEDP with MDF Bidding}
With uncertain wind generation and controllable loads in the day-ahead market, the market clearing process is to solve a CEDP with MDF bidding. The objective of the market clearing problem is to minimize the summation of the total \textit{expected} generation cost and the total payment for the MDF services to the load aggregators. Moreover, all the operational constraints for the charging/discharging power of the VBs are imposed in a probabilistic way, {\color{black}with the risk parameters for charging power and SOC as $\epsilon_i^f$ and $\epsilon_i^e$, respectively}. Denote the $t\times 1$ vector of the first $t$ random variables $\Omega_t$ from time slot $1$ to $t$ as $\widetilde{\boldsymbol{\Omega}}_t=(\Omega_1,\cdots,\Omega_t)^T$, which has zero mean and the covariance matrix is denoted as $\widetilde{\boldsymbol{\Sigma}}_t$. Then, the market clearing problem is formulated as the following chance-constrained program:

\begin{align}
& \underset{\boldsymbol{X}}{\text{min}}
& & \mathbb{E}\Big(\sum_{t\in \cal T}\sum_{i\in\mathcal{G}}C_i(P_{g,i,t})\Big)+\sum_{i\in\mathcal{F}}r_i(\boldsymbol{\alpha}_i) \\
& \text{s. t.}
& & \boldsymbol{1}_{n_g}^T\overline{\boldsymbol{P}}_{g,t}+\boldsymbol{1}_{n_w}^T\overline{\boldsymbol{P}}_{w,t}+\boldsymbol{1}_{n_f}^T\overline{\boldsymbol{P}}_{f,t}=\boldsymbol{1}_{n_d}^T\boldsymbol{P}_{d,t},  \label{mdf01:cst_1} \\
& & & 	\mathbb{P}_{\Omega_t}[P_{f,i,t}\leq \alpha_i^{R+}]\geq 1{-}\epsilon^f_i, \forall i \in \mathcal{F}, t \in [t_i^s,t_i^e],\label{mdf:cst_pf1}\\
& & & \mathbb{P}_{\Omega_t}[P_{f,i,t}\geq \alpha_i^{R-}]\geq 1{-}\epsilon^f_i, \forall i \in \mathcal{F},t \in [t_i^s,t_i^e],\label{mdf:cst_pf2}\\
& & & 	\mathbb{P}_{\widetilde{\boldsymbol{\Omega}}_t}[E_{i,t}\leq \alpha_i^{E+}]\geq 1{-}\epsilon^e_i, \forall i \in \mathcal{F},t \in [t_i^s,t_i^e],\label{mdf:cst_pf3}\\
& & & \mathbb{P}_{\widetilde{\boldsymbol{\Omega}}_t}[E_{i,t}\geq \alpha_i^{E-}]\geq 1{-}\epsilon^e_i, \forall i \in \mathcal{F},t \in [t_i^s,t_i^e],\label{mdf:cst_pf4}\\
& & & \textrm{constraints} \ (\ref{eq:affine_g1}),(\ref{mdf:ed_cst_g1}){-}(\ref{mdf:ed_cst_g2}), (\ref{mdf:ed_cst_l1}){-}(\ref{mdf:ed_cst_l2}), (\ref{mdf:alpha_1})-(\ref{mdf:lineflow_1}).\nonumber
\end{align}

In this optimization problem, the decision variables are 
$
	\boldsymbol{X}:=(\overline{\boldsymbol{P}}_{g,t},\overline{\boldsymbol{P}}_{f,t},{\boldsymbol{\beta}}_{g,t},{\boldsymbol{\beta}}_{f,t},\boldsymbol{\alpha},t\in \mathcal{T}),
$ where $\boldsymbol{\alpha} := \{\boldsymbol{\alpha}_i, i \in \cal F \}$.
Constraint \eqref{mdf01:cst_1} requires the expected power balance between supply and demand with controllable loads. Moreover, constraints (\ref{mdf:cst_pf1})--(\ref{mdf:cst_pf4}) are imposed such that the charging/discharging power and the SOC of VBs are maintained within the corresponding limits in a probabilistic manner.

The market clearing problem is hard to solve due to the fact that the objective function involves an expectation over uncertain wind generations. However, under the normality assumption for the forecast errors of the wind outputs, we can show that the market clearing problem can be equivalently reformulated into an SOCP, which can be solved optimally and efficiently.
\subsection{SOCP Reformulation of the Market Clearing Problem}
Assuming that the generation cost function is a quadratic function $
			C_i(P_{g,i,t})=c_{2,i}P^2_{g,i,t}+c_{1,i}P_{g,i,t}+c_{0,i},
$
we have
\begin{equation*}
		\mathbb{E}\big(C_i(P_{g,i,t})\big)=\mathbb{E}\big(c_{2,i}P^2_{g,i,t}\big)+\mathbb{E}\big(c_{1,i}P_{g,i,t}\big)+\mathbb{E}\big(c_{0,i}\big).
\end{equation*}
Also, because $\mathbb{E}(\Omega_t)=0$ and $\mathbb{E}(\Omega_t^2)=\textrm{Var}(\Omega_t)=\sigma_t^2$, from (\ref{eq:affine_g1}), we have $\mathbb{E}\big(c_{1,i}P_{g,i,t}\big)=c_{1,i}P_{g,i,t}$, $\mathbb{E}\big(c_{0,i}\big)=c_{0,i}$ and 
\begin{align*}
	 \mathbb{E}\big(c_{2,i}P^2_{g,i,t}\big)&=\mathbb{E}\big(c_{2,i}(\overline{P}^2_{g,i,t}{+}\Omega_t^2\beta_{g,i,t}^2-2\Omega_t\beta_{g,i,t})\big)\\
	&=\  c_{2,i}(\overline{P}^2_{g,i,t}{+}\sigma_t^2\beta_{g,i,t}^2).
\end{align*}
Therefore, the objective function can be expressed as
\begin{equation*}
\begin{aligned}
\sum_{t\in \cal T}\sum_{i\in\mathcal{G}}c_{2,i}(\overline{P}^2_{g,i,t}{+}\sigma_t^2\beta_{g,i,t}^2){+}c_{1,i}\overline{P}_{g,i,t}{+}c_{0,i}+\sum_{i\in\mathcal{F}}r_i(\boldsymbol{\alpha}_i).\label{mdf:obj_2}
\end{aligned}
\end{equation*}
Note that the objective function is in fact a convex (quadratic) function of the decision variables. 

Constraints (\ref{mdf:ed_cst_g1}){-}(\ref{mdf:ed_cst_g2}), (\ref{mdf:ed_cst_l1}){-}(\ref{mdf:ed_cst_l2}) and (\ref{mdf:cst_pf1})-(\ref{mdf:cst_pf4}) can be divided into three groups, depending on the random variables that they are related to. Specifically, constraints (\ref{mdf:ed_cst_g1}){-}(\ref{mdf:ed_cst_g2}) and (\ref{mdf:cst_pf1})--(\ref{mdf:cst_pf2}) can be converted into the following form: 
\begin{equation}
	\mathbb{P}_{\Omega_t}[A_{\Omega_t,i,t}(\boldsymbol{X})\cdot\Omega_t\leq b_{\Omega_t,i,t}(\boldsymbol{X})]\geq 1-\epsilon^{g/f}_i,\label{mdf:cc_type1}
\end{equation} 
where the probability is with respect to the random variable $\Omega_t$ and $\epsilon^{g/f}_i$ denotes either $\epsilon^{g}_i$ or $\epsilon^{f}_i$. Furthermore, $A_{\Omega_t,i,t}(\boldsymbol{X})$, which is a scalar, and $b_{\Omega_t,i,t}(\boldsymbol{X})$ are all linear expressions of the decision variables $\boldsymbol{X}$, as listed in Table \ref{mdf_table:1101}.

Similarly, constraints (\ref{mdf:cst_pf3})--(\ref{mdf:cst_pf4}) can be converted to:
\begin{equation}
\mathbb{P}_{\widetilde{\boldsymbol{\Omega}}_t}[A_{{\widetilde{\boldsymbol{\Omega}}_t},i,t}(\boldsymbol{X})\cdot{\widetilde{\boldsymbol{\Omega}}_t}\leq b_{{\widetilde{\boldsymbol{\Omega}}_t},i,t}(\boldsymbol{X})]\geq 1-\epsilon^e_{i},\label{mdf:cc_type2}
\end{equation}
and constraints (\ref{mdf:ed_cst_l1}){-}(\ref{mdf:ed_cst_l2}) can be converted to:
\begin{equation}
\mathbb{P}_{\boldsymbol{\omega}_t}[A_{{\boldsymbol{\omega}_t},l,t}(\boldsymbol{X})\cdot{\boldsymbol{\omega}_t}\leq b_{{\boldsymbol{\omega}_t},l,t}(\boldsymbol{X})]\geq 1-\epsilon_{l}.\label{mdf:cc_type3}
\end{equation}
In the above constraints, $A_{{\widetilde{\boldsymbol{\Omega}}_t},i,t}(\boldsymbol{X})$ and  $A_{{\boldsymbol{\omega}_t},l,t}(\boldsymbol{X})$ are row vectors whose elements are linear in $\boldsymbol{X}$. Besides, $b_{{\widetilde{\boldsymbol{\Omega}}_t},i,t}(\boldsymbol{X})$ and  $b_{{\boldsymbol{\omega}_t},l,t}(\boldsymbol{X})$ are also linear in $\boldsymbol{X}$. Specifically, for constraint (\ref{mdf:ed_cst_l1}), 
\begin{equation}
\begin{aligned}
 A_{{\boldsymbol{\omega}_t},l,t}(\boldsymbol{X})=& {\color{black}e_{l}^T}\cdot\boldsymbol{\Gamma}(-H_g\boldsymbol{\beta}_{g,t}\boldsymbol{1}_{n_w}^T+H_w-H_f\boldsymbol{\beta}_{f,t}\boldsymbol{1}_{n_w}^T), \\
 b_{{\boldsymbol{\omega}_t},l,t}(\boldsymbol{X})=& F_{l}^{max}-{\color{black}e_{l}^T}\cdot\boldsymbol{\Gamma}(H_g\overline{\boldsymbol{P}}_{g,t}+H_w\overline{\boldsymbol{P}}_{w,t}\\
 &+H_f\overline{\boldsymbol{P}}_{f,t}-H_d\boldsymbol{P}_{d,t}),
\end{aligned}
\end{equation} 
and for constraint (\ref{mdf:ed_cst_l2}),
\begin{equation}
\begin{aligned}
A_{{\boldsymbol{\omega}_t},l,t}(\boldsymbol{X})=&- {\color{black}e_{l}^T}\cdot\boldsymbol{\Gamma}(-H_g\boldsymbol{\beta}_{g,t}\boldsymbol{1}_{n_w}^T+H_w-H_f\boldsymbol{\beta}_{f,t}\boldsymbol{1}_{n_w}^T), \\
b_{{\boldsymbol{\omega}_t},l,t}(\boldsymbol{X})=& F_{l}^{max}+{\color{black}e_{l}^T}\cdot\boldsymbol{\Gamma}(H_g\overline{\boldsymbol{P}}_{g,t}+H_w\overline{\boldsymbol{P}}_{w,t}\\
&+H_f\overline{\boldsymbol{P}}_{f,t}-H_d\boldsymbol{P}_{d,t}),
\end{aligned}
\end{equation} 
where $e_l$ is a $L \times 1$ vector with all the elements equal to zero except the $l$-th element, which is equal to one.

Since the chance constraints are all linear in both the decision variables and random variables, under the normality assumption for the deviation of wind turbine outputs, they can be equivalently reformulated into second order cone constraints \cite{bienstock2014chance}. For example, as for constraint (\ref{mdf:cc_type3}), since $\boldsymbol{\omega}_t \sim \mathcal{N}(\boldsymbol{0},\Sigma_t)$, $A_{{\boldsymbol{\omega}_t},l,t}(\boldsymbol{X})\cdot{\boldsymbol{\omega}_t}- b_{{\boldsymbol{\omega}_t},l,t}(\boldsymbol{X}) \sim \mathcal{N}(\boldsymbol{0},A_{{\boldsymbol{\omega}_t},l,t}(\boldsymbol{X})\Sigma_t(A_{{\boldsymbol{\omega}_t},l,t}(\boldsymbol{X}))^T)$, then
\begin{align*}
&	\mathbb{P}_{\boldsymbol{\omega}_t}[A_{{\boldsymbol{\omega}_t},l,t}(\boldsymbol{X})\cdot{\boldsymbol{\omega}_t}\leq b_{{\boldsymbol{\omega}_t},l,t}(\boldsymbol{X})] \\
=\ & \mathbb{P}_{\boldsymbol{\omega}_t}[A_{{\boldsymbol{\omega}_t},l,t}(\boldsymbol{X})\cdot{\boldsymbol{\omega}_t}-b_{{\boldsymbol{\omega}_t},l,t}(\boldsymbol{X})\le 0] \\
=\ &\mathbb{P}_{Z}[Z\leq b_{{\boldsymbol{\omega}_t},l,t}(\boldsymbol{X})/\sqrt{A_{{\boldsymbol{\omega}_t},l,t}(\boldsymbol{X})\Sigma_t(A_{{\boldsymbol{\omega}_t},l,t}(\boldsymbol{X}))^T}],\\
=\ & \Phi\Big(b_{{\boldsymbol{\omega}_t},l,t}(\boldsymbol{X})/||\Sigma_t^{1/2}(A_{{\boldsymbol{\omega}_t},l,t}(\boldsymbol{X}))^T||_2\Big),
\end{align*} 
where $Z$ is a standard normal variable and $\Phi(\cdot)$ is its cumulative distribution function (CDF). Therefore, we have
\begin{align}
	&\textrm{constraint} \ (\ref{mdf:cc_type3}) \nonumber\\
 \Leftrightarrow \	& \Phi\Big(b_{{\boldsymbol{\omega}_t},l,t}(\boldsymbol{X})/||\Sigma_t^{1/2}(A_{{\boldsymbol{\omega}_t},l,t}(\boldsymbol{X}))^T||_2\Big) \ge 1-\epsilon_l \nonumber\\
 \Leftrightarrow \ & \Phi^{-1}(1-\epsilon_{l})||\Sigma_t^{1/2}(A_{{\boldsymbol{\omega}_t},l,t}(\boldsymbol{X}))^T||_2 \leq b_{\boldsymbol{\omega}_t,l,t}(\boldsymbol{X}), \label{mdf:socp_3}
\end{align}
where $\Phi^{-1}(\cdot)$ is the inverse cumulative distribution function of the standard normal distribution. Similarly, constraints (\ref{mdf:cc_type2}) and (\ref{mdf:cc_type1}) are equivalent to the following two constraints:
\begin{align}
&\Phi^{-1}(1-\epsilon_{i}^e)||\widetilde{\boldsymbol{\Sigma}}_t^{1/2}(A_{{\widetilde{\boldsymbol{\Omega}}_t},i,t}(\boldsymbol{X}))^T||_2 \leq b_{\widetilde{\boldsymbol{\Omega}}_t,i,t}(\boldsymbol{X}),  \label{mdf:socp_1}\\
	& \Phi^{-1}(1-\epsilon_i^{g/f})|\sigma_t A_{\Omega_t,i,t}(\boldsymbol{X})|\leq b_{\Omega_t,i,t}(\boldsymbol{X}). \label{mdf:socp_2}
\end{align}
Note that constraints (\ref{mdf:socp_3})-(\ref{mdf:socp_2}) are SOC constraints. Consequently, we have successfully reformulated the market clearing problem into an SOCP which can be efficiently solved by commercial solvers such as CVX \cite{grant2008cvx}.
\begin{table}[t!]
	\caption{Expressions of $A$ and $b$ for constraints (\ref{mdf:ed_cst_g1}),(\ref{mdf:ed_cst_g2}), (\ref{mdf:cst_pf1})-(\ref{mdf:cst_pf4})}
	\label{mdf_table:1101}
	\centering
	\begin{tabular}{ccc}
		\hline
		& 	$A_{\Omega_t/\widetilde{\boldsymbol{\Omega}}_t,i,t}(\boldsymbol{X})$ & $b_{\Omega_t/\widetilde{\boldsymbol{\Omega}}_t,i,t}(\boldsymbol{X})$ \\
		\hline
		(\ref{mdf:ed_cst_g1}) & $-\beta_{g,i,t}$ & $P^{max}_{g,i}-\overline{P}_{g,i,t}$\\
		\hline
		(\ref{mdf:ed_cst_g2}) & $\beta_{g,i,t}$ & $\overline{P}_{g,i,t}-P^{min}_{g,i}$ \\
		\hline 
		(\ref{mdf:cst_pf1}) & $-\beta_{f,i,t}$ & $\alpha^{R+}_{i}-\overline{P}_{g,i,t}$\\
		\hline 
		(\ref{mdf:cst_pf2}) & $\beta_{f,i,t}$ & $\overline{P}_{g,i,t}-\alpha^{R-}_{i}$\\
		\hline 
		(\ref{mdf:cst_pf3}) & $(\beta_{f,i,1},\beta_{f,i,2},\ldots,\beta_{f,i,t})$ & $\alpha_i^{E+}+\sum_{\tau=1}^{t}\overline{P}_{f,i,\tau}$\\
		\hline
		(\ref{mdf:cst_pf4}) & $-(\beta_{f,i,1},\beta_{f,i,2},\ldots,\beta_{f,i,t})$ & $-\alpha_i^{E-}-\sum_{\tau=1}^{t}\overline{P}_{f,i,\tau}$\\
		\hline
	\end{tabular}
	
\end{table}
\section{Case Studies}\label{mdf:sec4}
\subsection{Improving LMP Profiles by MDF Bidding}
First, we use a six-bus transmission network \cite{kargarian2014system} as shown in Fig. \ref{mdf_fig:121} to implement our day-ahead MDF market when the wind power outputs follow the mean values, i.e., $\Sigma_t$ is a zero matrix for any $t\in \mathcal{T}$. In this case, the left-hand sides of constraints \eqref{mdf:socp_3}-\eqref{mdf:socp_2} become zero and thus, the original chance constraints \eqref{mdf:ed_cst_g1}{-}\eqref{mdf:ed_cst_g2}, \eqref{mdf:ed_cst_l1}{-}\eqref{mdf:ed_cst_l2} and \eqref{mdf:cst_pf1}-\eqref{mdf:cst_pf4} are reduced to traditional deterministic hard constraints. Therefore, we can study the impact of the MDF bidding on the LMP profiles in the deterministic settings.
 
\begin{figure}[!t]
	\centering
	\includegraphics[width=3.0in]{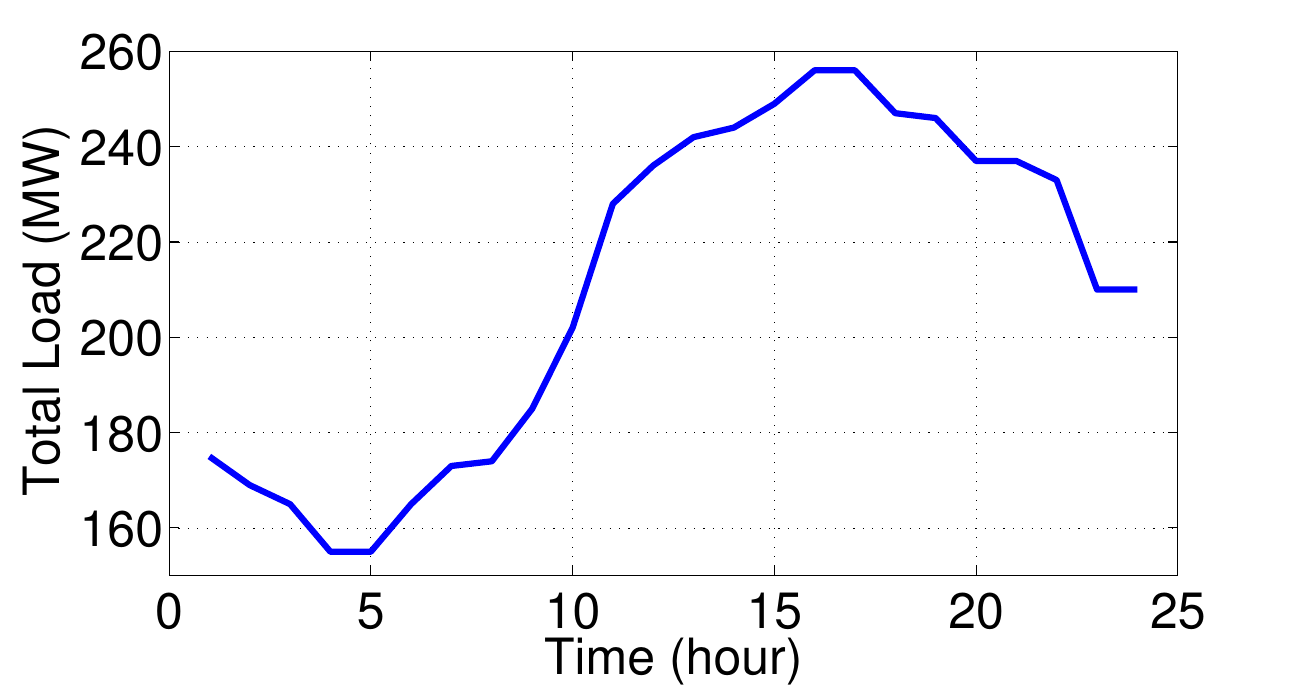}
	\caption{Total hourly net load over 24-hour horizon.}
	\label{mdf_fig:135}
\end{figure}

The network information of the six-bus system is listed in Table \ref{mdf_table:122}, where F.B. and T.B. stand for from-bus and to-bus respectively. Also, X is the line reactance in p.u. and F.L. denotes the flow limits in MW. Buses 1, 2, and 6 are generator buses, and their generation cost functions and operation limits are shown in Table \ref{mdf_table:121}. The other three buses are load buses, in each of which there is a load aggregator participating in the MDF market. Since the wind power follows the mean value, we are only interested in the net load. The total net load of all the buses over 24 hours is given in Fig. \ref{mdf_fig:135}. Load buses 3, 4, and 5 consume 20\%, 40\% and 40\% of the total net load, respectively. The power base is 100MVA. For the MDF bids, we assume that all the three load aggregators have the same bid for the VBs (in p.u.), i.e., $R_i^{min}=-0.1,R_i^{max}=0.3,E_i^{min}=-0.5,E_i^{max}=0.3, i=3,4,5$. However, the available service periods are different, which are [13:00,19:00], [09:00,16:00] and [16:00,23:00] for buses 3, 4 and 5, respectively. Next, we study the LMP profiles of the load buses in the following three cases:
\begin{figure}[!t]
	\centering
	\includegraphics[width=3.0in]{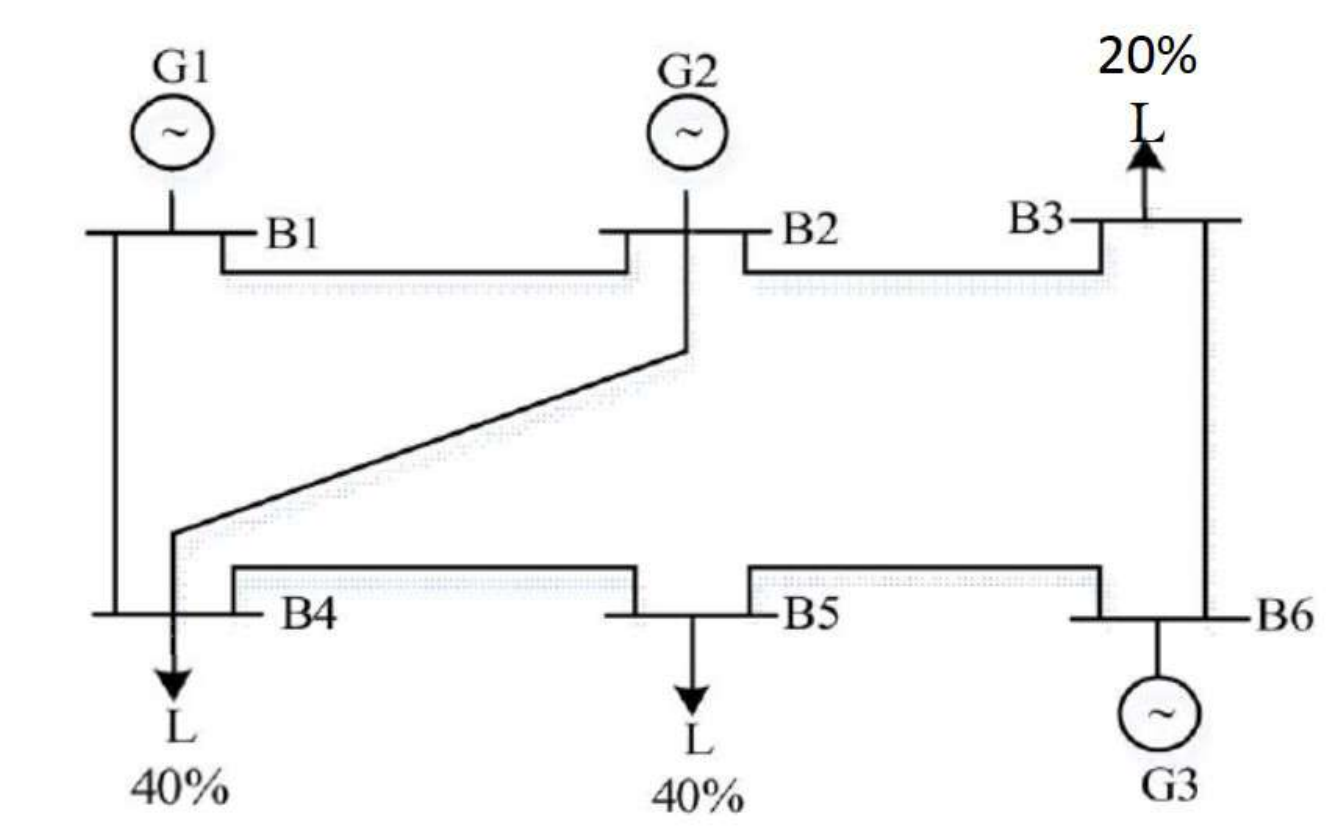}
	\caption{Six-bus transmission network.}
	\label{mdf_fig:121}
\end{figure}
\begin{table}[!t]
	\caption{Network Information}
	\label{mdf_table:122}
	\centering
	\begin{tabular}{cccccccc}
		\hline
		F.B. & T.B. & X  &   F.L. & F.B. & T.B. & X & F.L.\\
		\hline
		1 & 2 & 0.170 & 60 & 3 & 6 & 0.018 &  180  \\
		\hline
		1 & 4 & 0.258 &  70 & 4 & 5 &  0.037 & 190   \\
		\hline
		2 & 3  & 0.037 & 190  & 5 & 6 & 0.140 & 180   \\
		\hline
		2 & 4 & 0.197 &  200  & & & & \\
		\hline
	\end{tabular}
\end{table}
\begin{table}[!t]
	\caption{Generator Data}
	\label{mdf_table:121}
	\centering
	\begin{tabular}{ccccccc}
		\hline
		Index & $P_G^{min} $(MW) & $P_G^{max} $(MW) &   a(\$/$\textrm{MW}^2\textrm{h}$) & b(\$/MWh) & c(\$) \\
		\hline
		G1 & 40 & 220 &  0.03 & 7 & 100   \\
		\hline
		G2 & 10 & 200 &  0.07 & 10 & 104 \\
		\hline
		G3 & 0  & 25 & 0.05 & 8 & 110    \\
		\hline
	\end{tabular}
\end{table}
   \begin{table} [!t]
   	\centering
   	\caption{MDF Clearing Results}
   	\label{mdf_multiprogram}
   	\begin{tabular}{ccc|cc}
   		\hline
   		&  \multicolumn{2}{c|}{$\gamma_i^P=50,\gamma_i^E=50$} & \multicolumn{2}{c}{$\gamma_i^P=500,\gamma_i^E=500$} \\
   		\hline
   		&  $\alpha_i^{R+}$ & $\alpha_i^{E-}$ & $\alpha_i^{R+}$ & $\alpha_i^{E-}$ \\
   		\hline
   		B3 & 0.129 & -0.5 & 0.095  & -0.5 \\
   		\hline
   		B4 & 0.179  & -0.5 & 0.134 & -0.5 \\
   		\hline
   		B5 &  0.156 & -0.5 & 0.085 & -0.5 \\
   		\hline   		   		
   	\end{tabular}
   \end{table}
\begin{itemize}
	\item Benchmark case: traditional ED problem without MDF bidding.
	\item Low-reward MDF case: load aggregators offer cheap MDF with $\gamma_i^P=50, \gamma^E_i=50,i=3,4,5.$
	\item High-reward MDF case: load aggregators offer expensive MDF with $\gamma_i^P=500, \gamma^E_i=500,i=3,4,5.$
\end{itemize}
\begin{figure*}[t!]
	\centering     
	\subfigure[Benchmark case.]{\label{mdf_fig:2a1}\includegraphics[width=55mm]{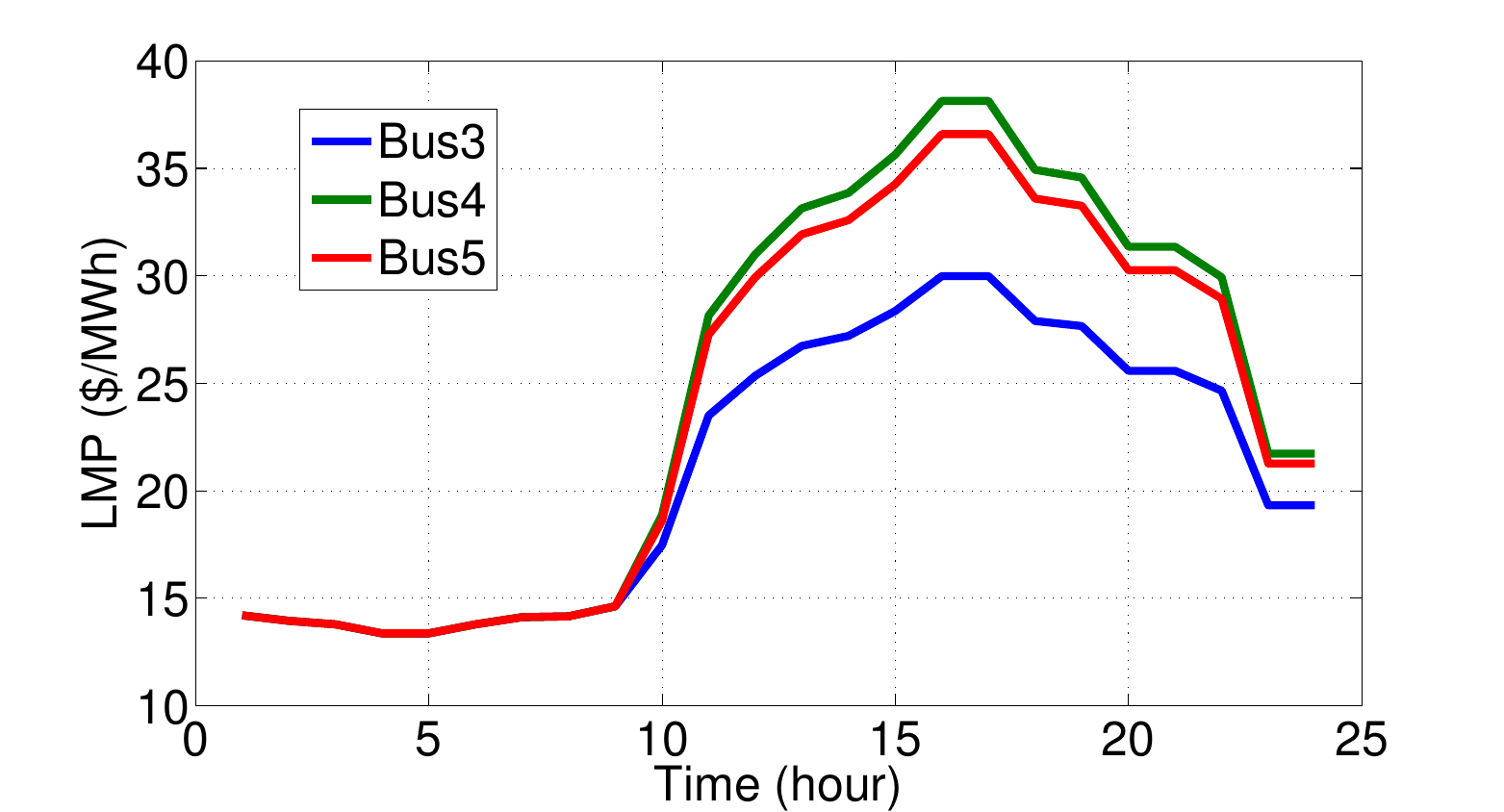}}
	\subfigure[Low-reward MDF case.]{\label{mdf_fig:2b1}\includegraphics[width=55mm]{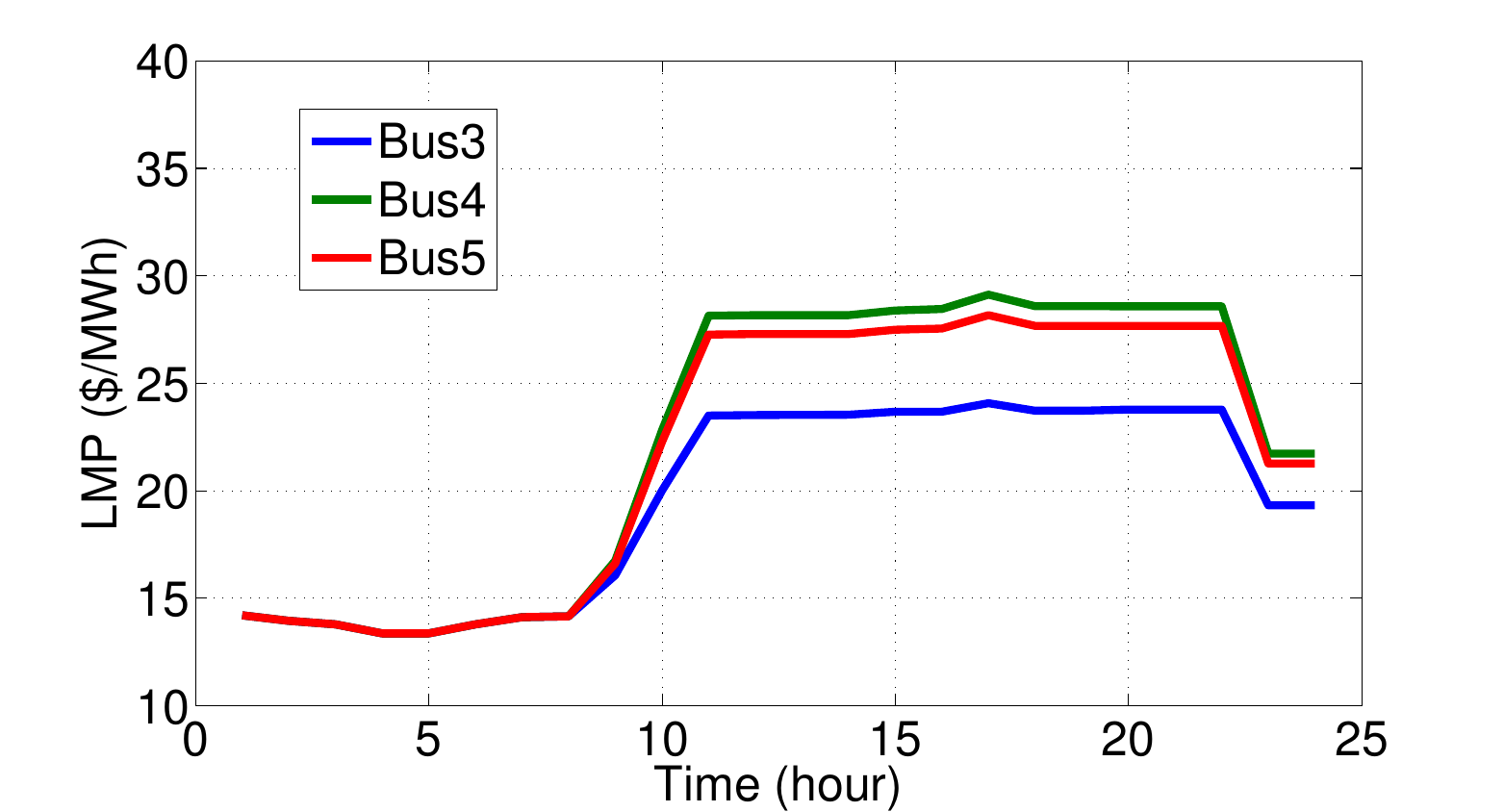}}
	\subfigure[High-reward MDF case.]{\label{mdf_fig:2c1}\includegraphics[width=55mm]{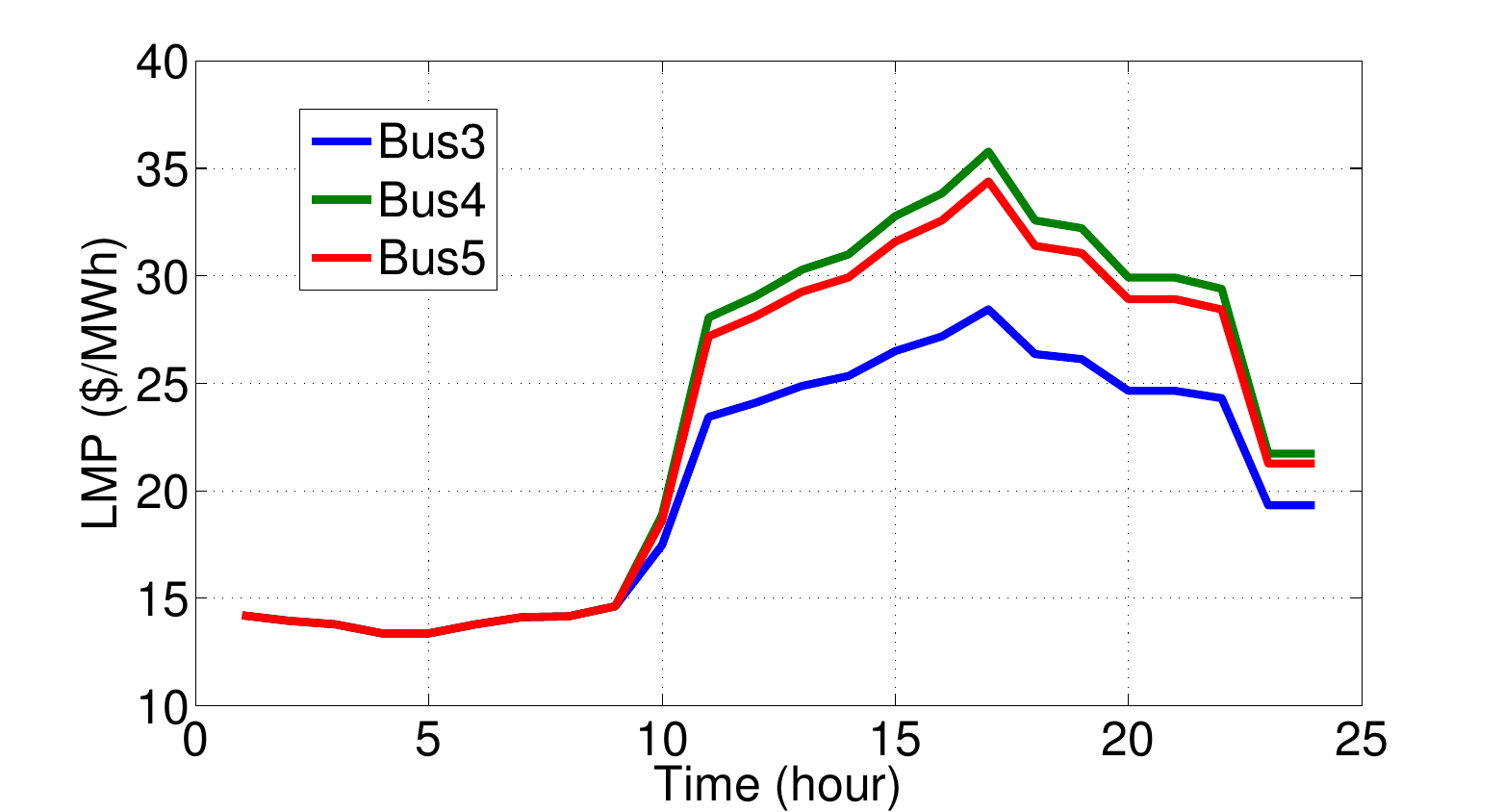}}
	\caption{LMPs for load buses in three cases.}\label{mdf_fig:212}
\end{figure*} 

By solving the market clearing problem in the three cases, we obtain and compare the hourly LMPs at buses 3, 4, and 5 in Fig. \ref{mdf_fig:212}, and show the cleared MDF bids in Table \ref{mdf_multiprogram}. Note that the values of $\alpha_i^{R-}$ and $\alpha_i^{E+}$ are always equal to zero in all three cases. This is because non-zero values of $\alpha_i^{R-}$ and $\alpha_i^{E+}$ mean that the system needs the load aggregators to increase their consumption of loads, which cannot bring any benefit to the system. From the benchmark case in Fig. \ref{mdf_fig:2a1}, we can observe that the LMPs for the load buses vary a lot between 10:00 to 23:00 due to the large variation and the high peak-to-average ratio of the net load. After integrating the MDF bids into the market-clearing problem, it can be seen from Fig. \ref{mdf_fig:2b1} and Fig.\ref{mdf_fig:2c1} that the average values of the LMP profiles decrease and their variations are reduced. It can be observed that those improvements in the LMP profiles come from both the energy and power flexibilities in the MDF bids by comparing the low-reward and high-reward MDF cases in Table \ref{mdf_multiprogram}.
Note that in both cases, the accepted values of $\alpha_i^{E-}$ for all the three buses always reach their maximum values (i.e., -0.5). This is because the energy flexibility offered by the MDF bids is much cheaper than the generation cost, and hence is fully accepted to minimize the total cost. It can also be observed that the energy flexibility mainly reduces the average values of LMP profiles in both cases. However, the variations of the LMP profiles can still be quite different with the same energy flexibility. Furthermore, the accepted power limits $\alpha_i^{R+}$ in the high-reward MDF case decrease by 26.4\%, 25.2\% and 45.5\% for buses 3, 4 and 5, compared to those in the low-reward MDF case. Therefore, the decrease of $\alpha_i^{R+}$ increases the variations of the LMPs significantly by comparing Fig. \ref{mdf_fig:2b1} and Fig. \ref{mdf_fig:2c1}.
Thus, from this case study, we understand that the energy flexibility and power flexibility reshape the LMP profiles from different aspects: the energy flexibility helps reduce the average LMP values and the power flexibility is important for flattening the LMPs. In addition, the reward parameters submitted in the MDFs can greatly influence the LMP profiles. Although it is out of the scope of this paper to study how to choose the MDF bid from the individual aggregator's perspective or how to guarantee each aggregator submits its truthful bids from the market operator's perspective, it is a very interesting and important problem for future research.

\subsection{Nine-bus System}
\begin{figure}[!t]
	\centering
	\includegraphics[width=3.5in]{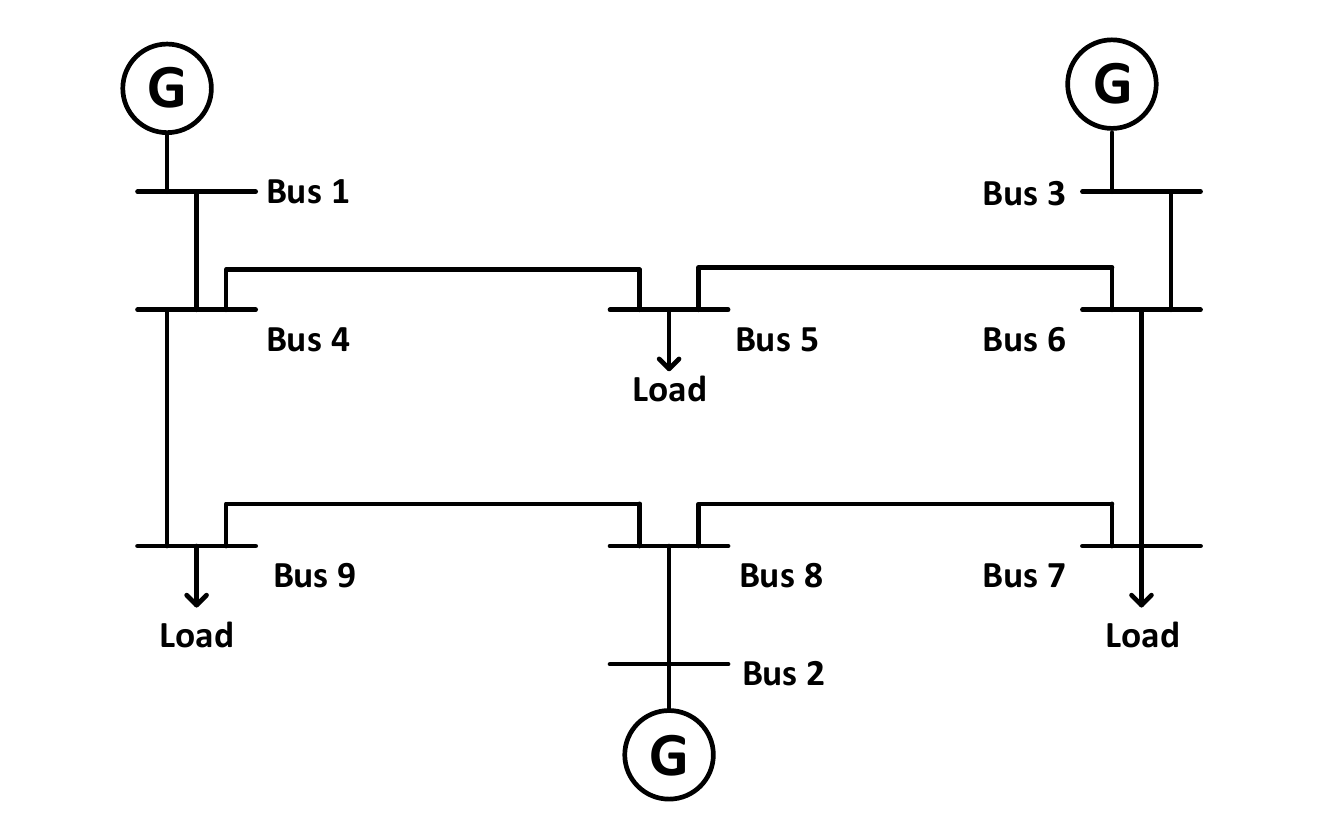}
	\caption{Nine-bus transmission network.}
	\label{mdf_fig:9bus}
\end{figure}

\begin{figure}[t!]
	\centering     
	\subfigure[Base load profiles in p.u.]{\label{mdf_fig:4a1}\includegraphics[width=42mm]{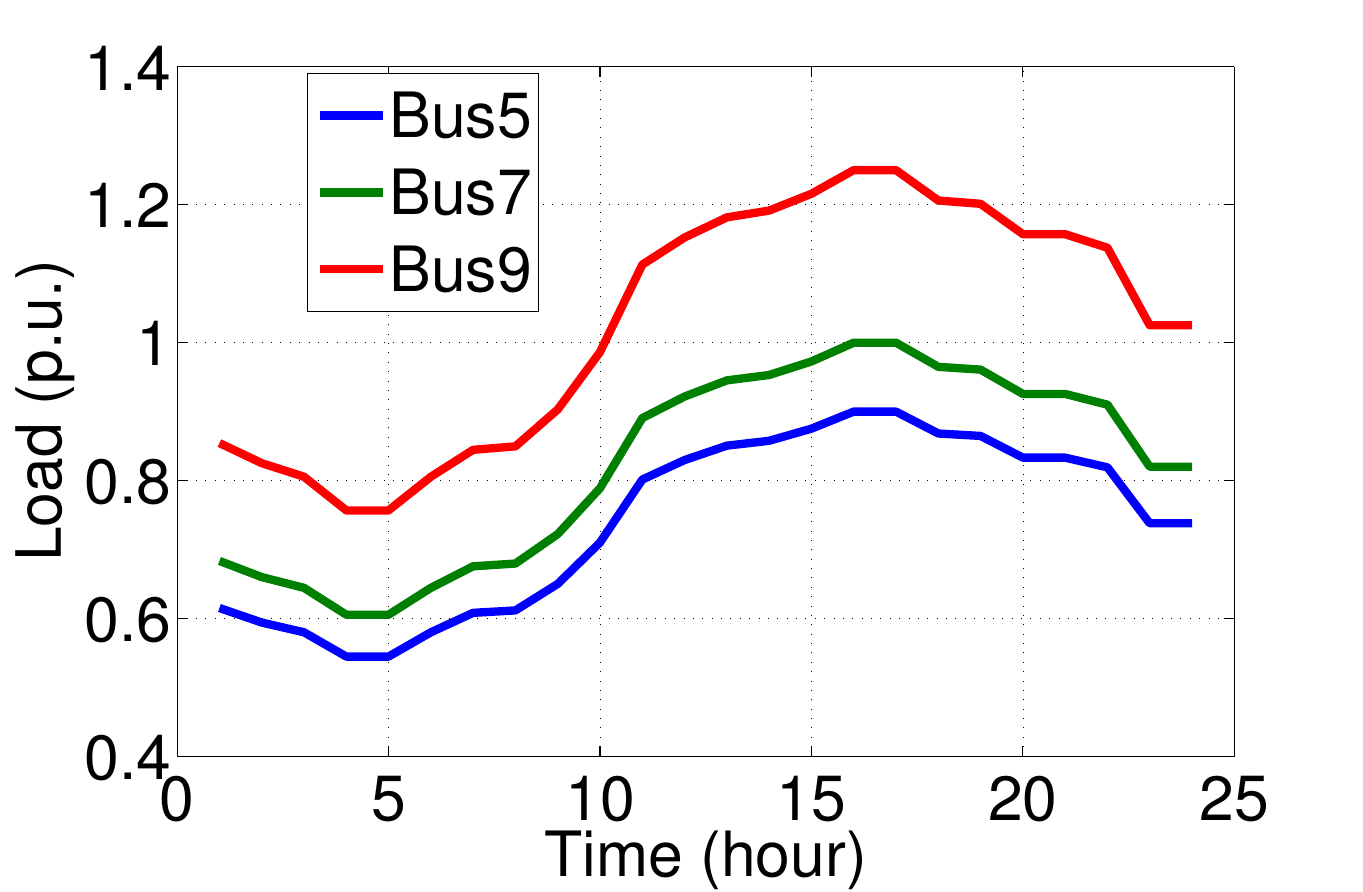}}
	\subfigure[Wind power output mean in p.u.]{\label{mdf_fig:4b1}\includegraphics[width=42mm]{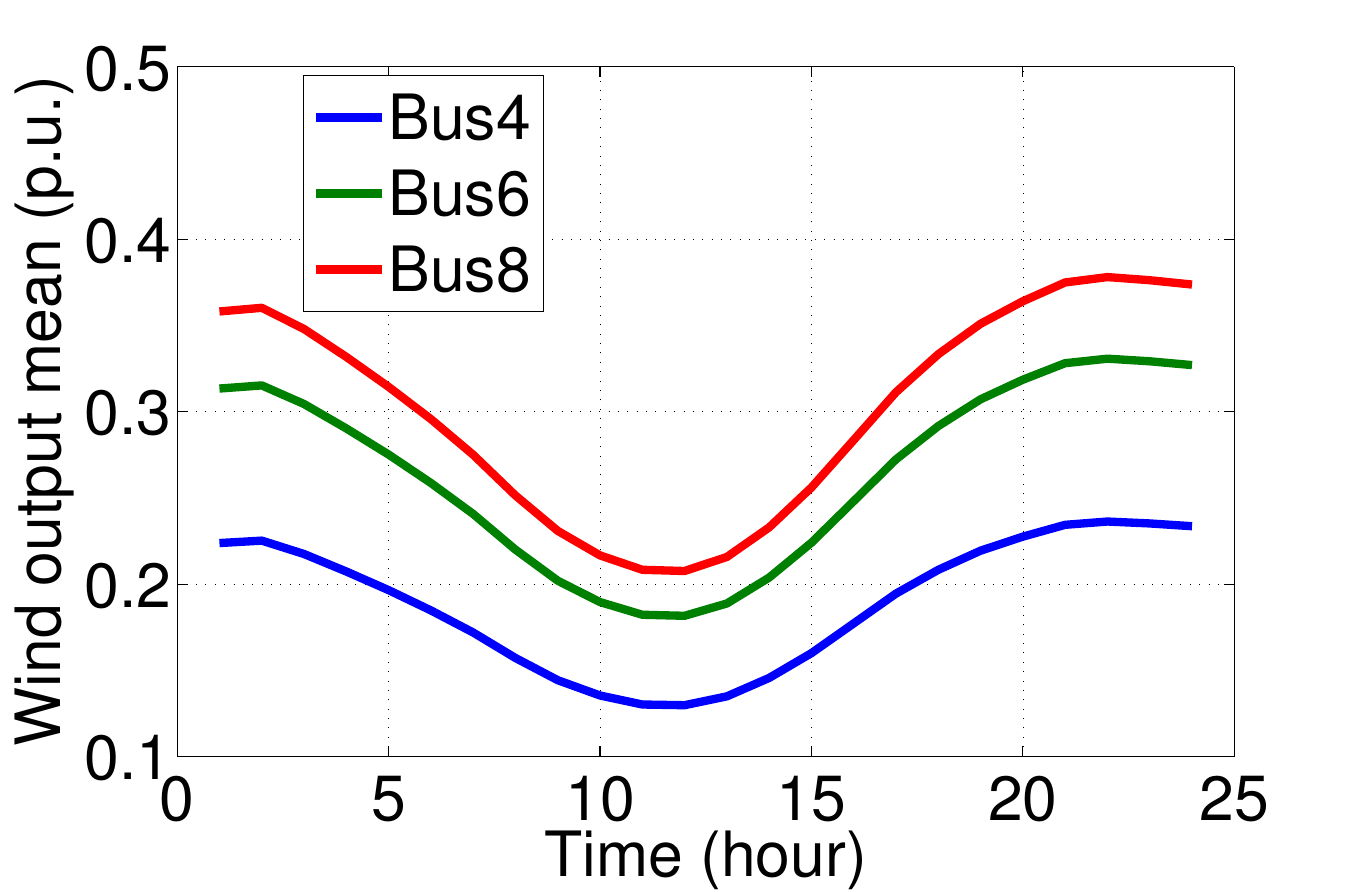}}
			\caption{Base load profiles and wind power output mean in p.u.}\label{mdf_fig:444}
\end{figure}

\begin{figure*}[t!]
	\vspace{-0.5cm}
		\centering     
		\subfigure[Cleared $|\alpha^{E-}|$ at bus 5.]{\label{mdf_fig:3a1}\includegraphics[width=59mm]{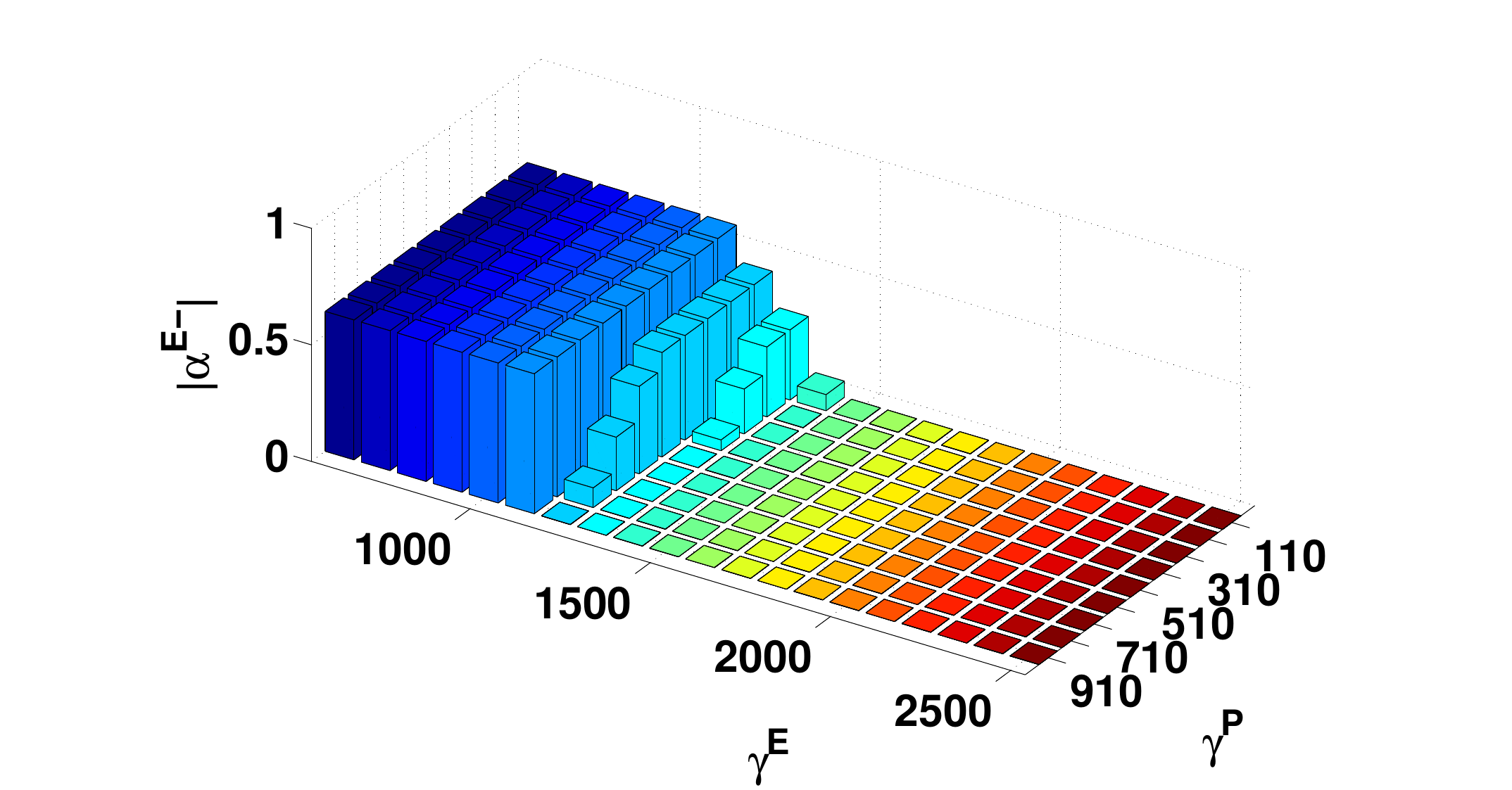}}
		\subfigure[Cleared $|\alpha^{E-}|$ at bus 7.]{\label{mdf_fig:3b1}\includegraphics[width=59mm]{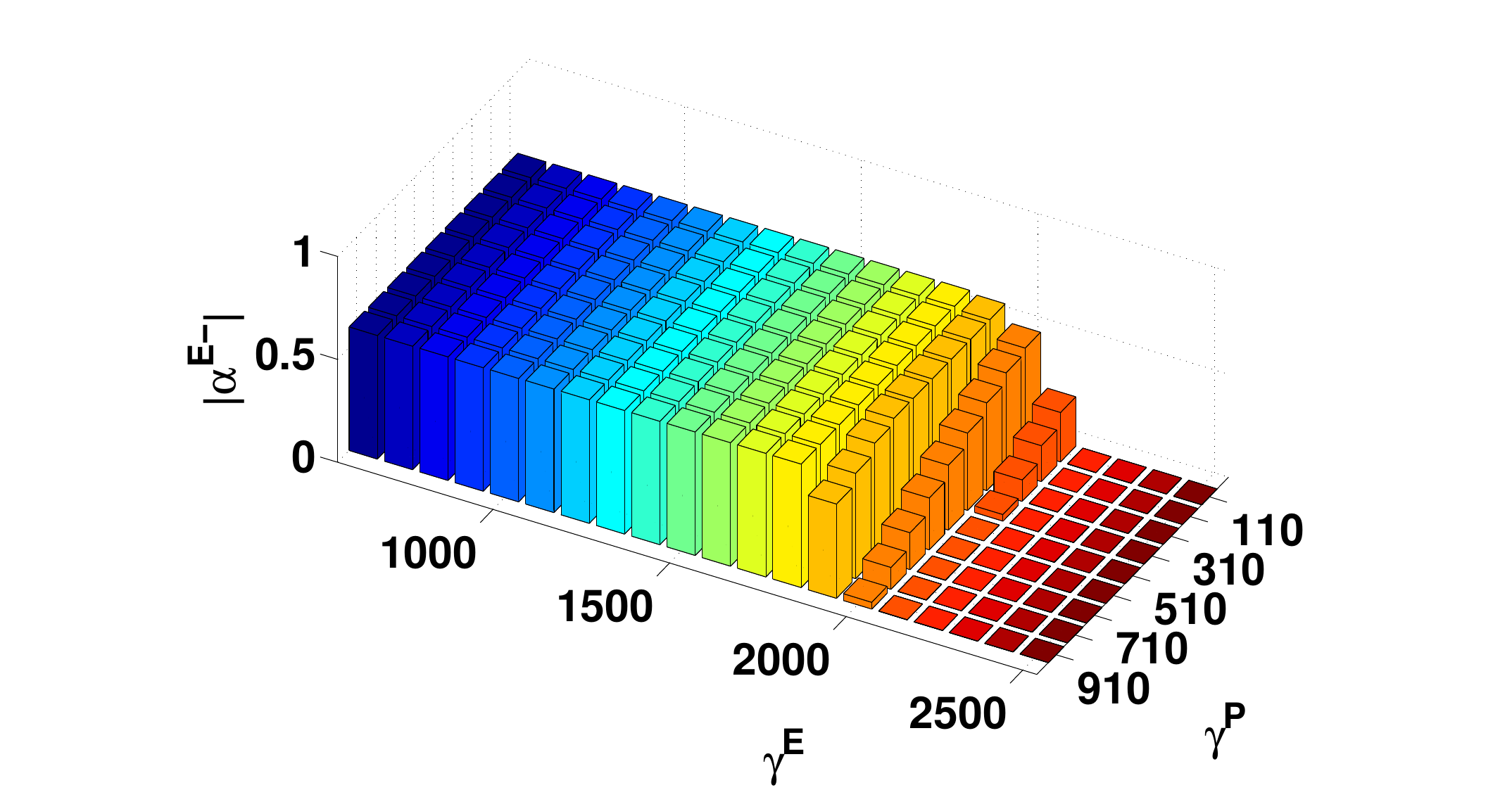}}
		\subfigure[Cleared $|\alpha^{E-}|$ at bus 9.]{\label{mdf_fig:3c1}\includegraphics[width=59mm]{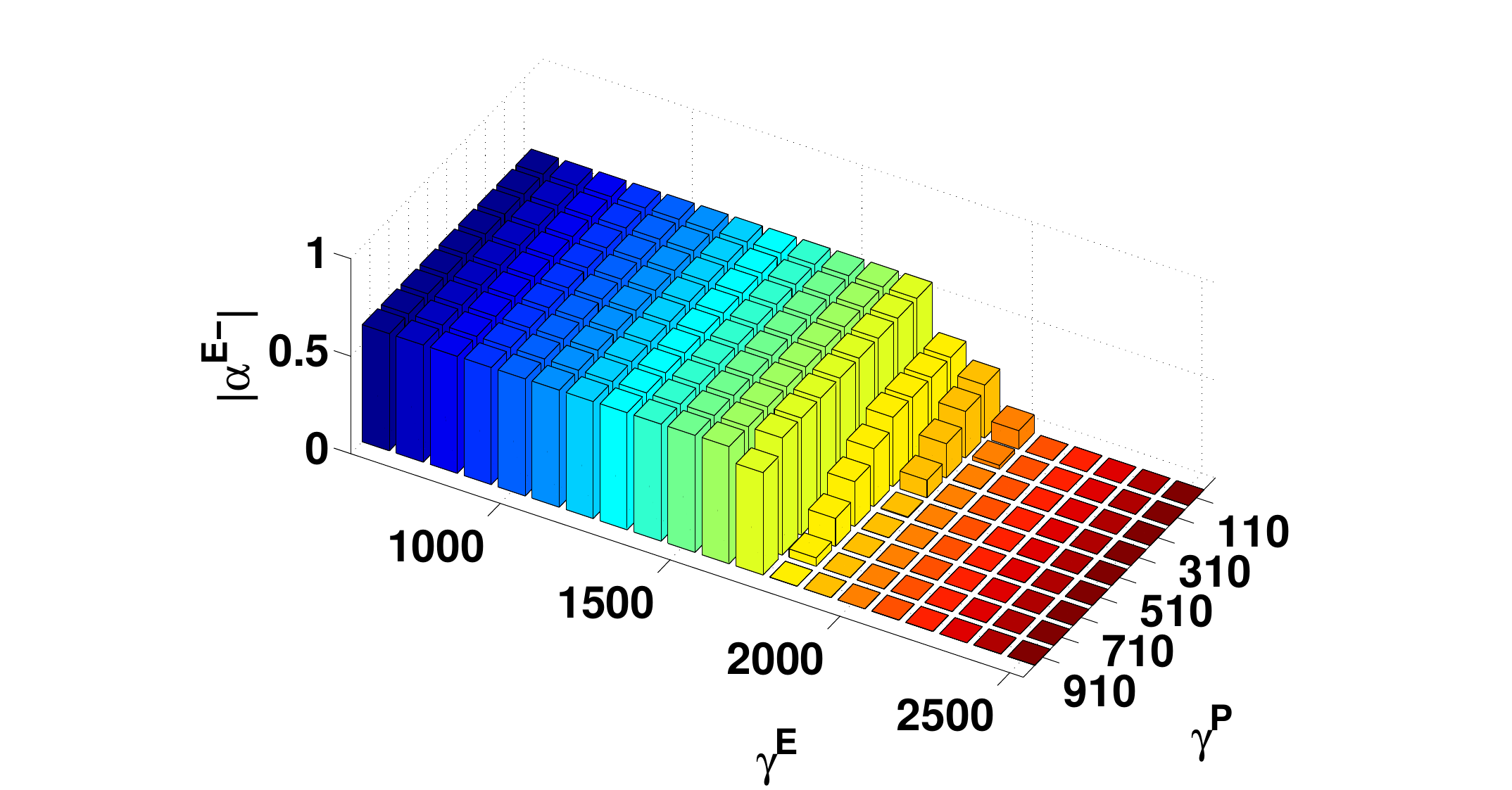}}
		\subfigure[Cleared $|\alpha^{R+}|$ at bus 5.]{\label{mdf_fig:3a2}\includegraphics[width=59mm]{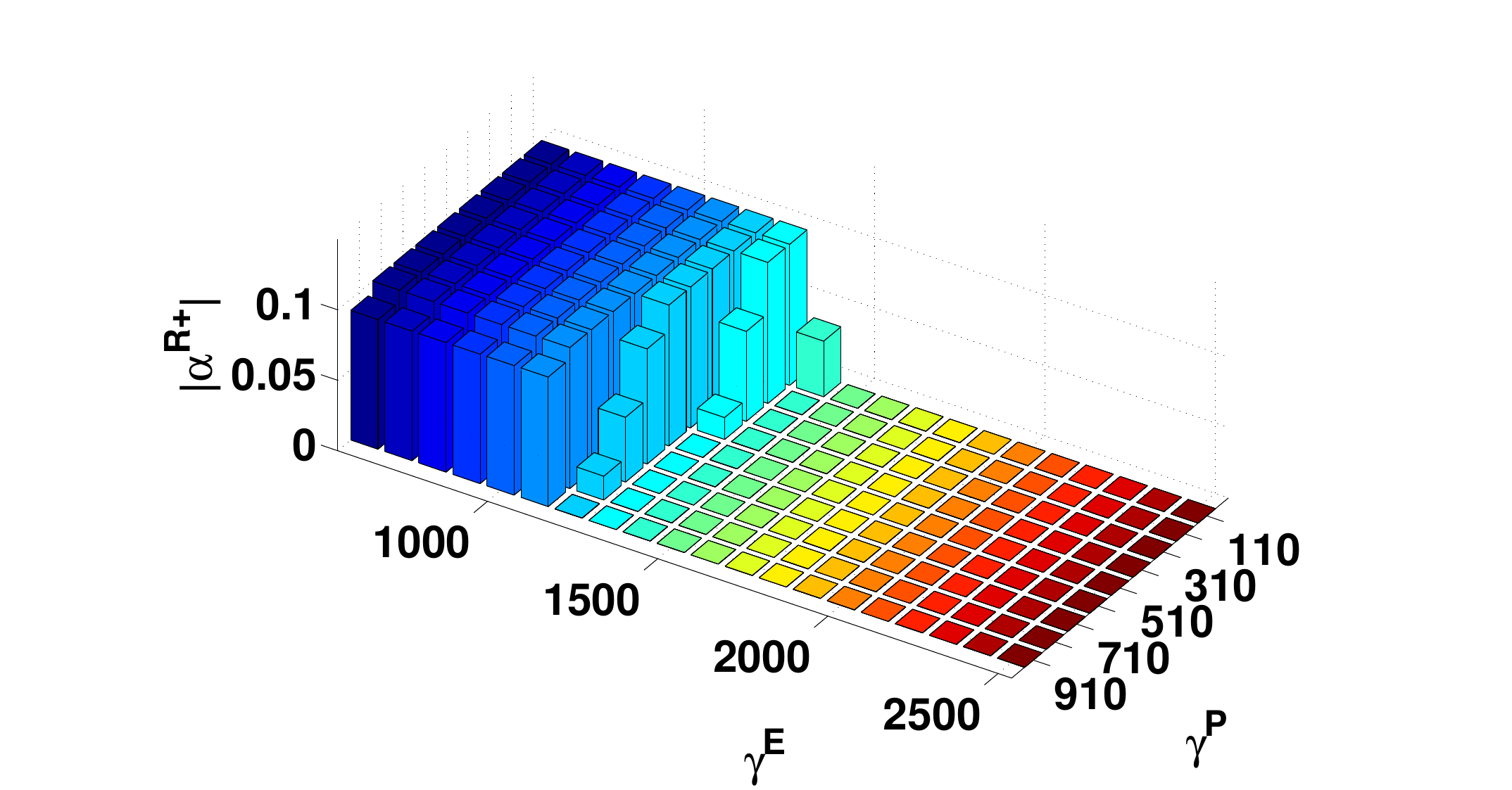}}
		\subfigure[Cleared $|\alpha^{R+}|$ at bus 7.]{\label{mdf_fig:3b2}\includegraphics[width=59mm]{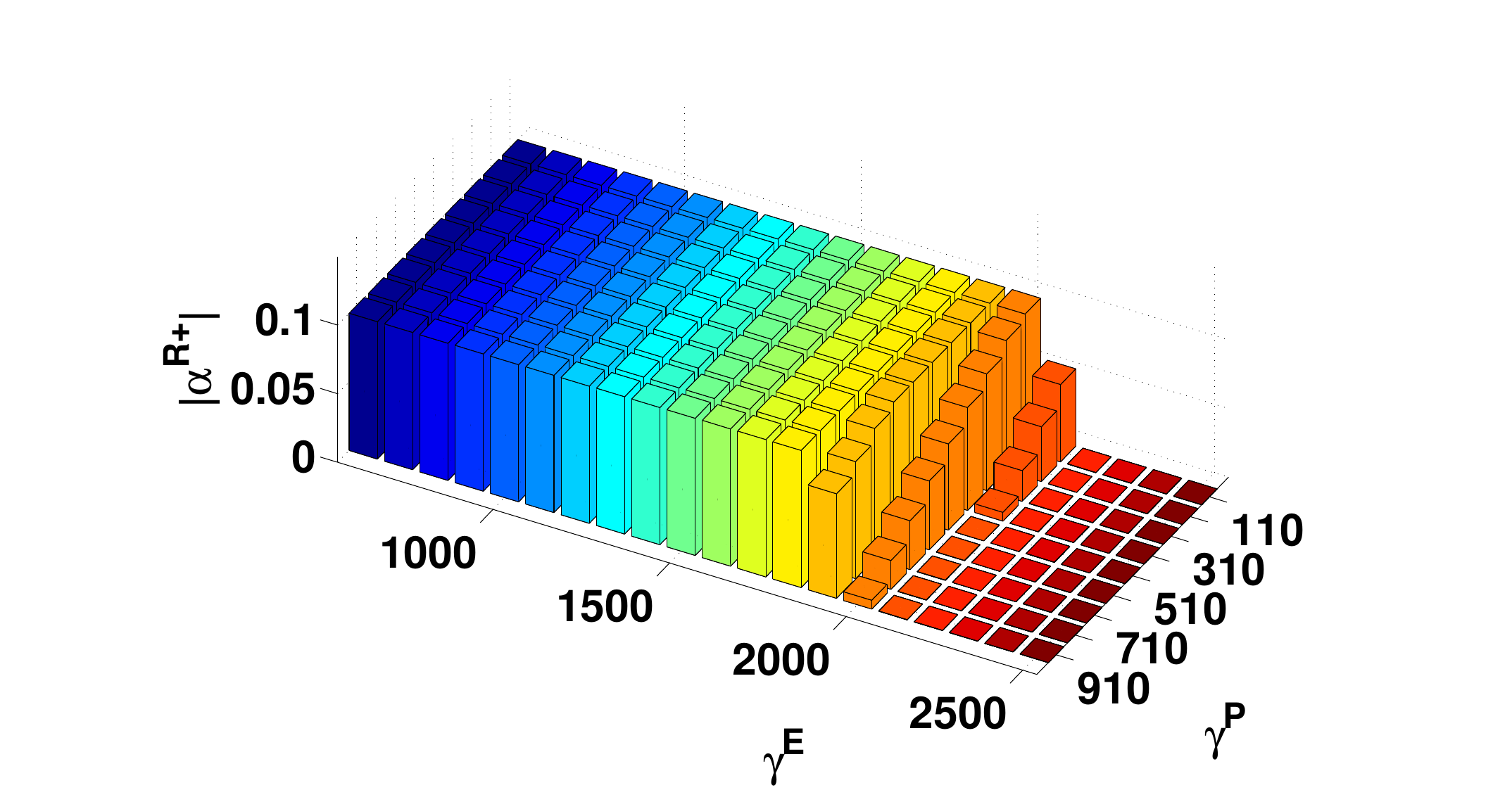}}
		\subfigure[Cleared $|\alpha^{R+}|$ at bus 9.]{\label{mdf_fig:3c2}\includegraphics[width=59mm]{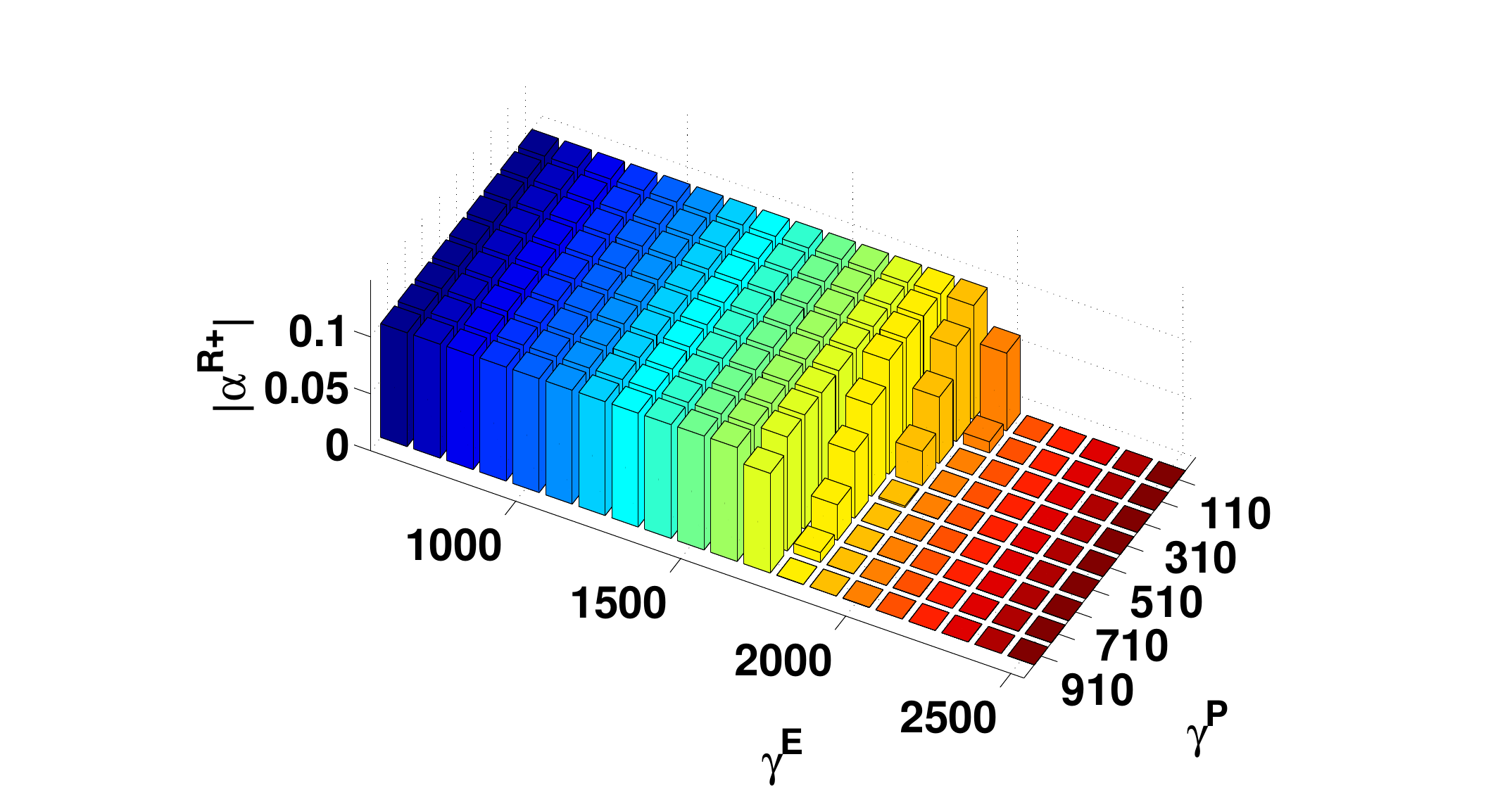}}
		\caption{Market clearing results for $\gamma^P$ ranging from 110 to 910 and $\gamma^E$ ranging from 100 to 2500.}\label{mdf_fig:333}
\end{figure*}

In order to show the effectiveness of the MDF bidding under uncertain wind generation, we apply our proposed MDF market to a modified IEEE 9-bus system shown in Fig. \ref{mdf_fig:9bus} \cite{zimmerman2011matpower}. There are three dispatchable generators located at buses 1, 2, and 3, respectively. In addition, there are three load aggregators at buses 5, 7 and 9, respectively and their base load profiles (in p.u.) are shown in Fig. \ref{mdf_fig:4a1}. Moreover, at buses 4, 6, and 8, each bus has a wind turbine with the mean of their outputs shown in Fig. \ref{mdf_fig:4b1}. The risk parameters are $\epsilon^g_i=\epsilon^f_i=\epsilon^e_i=0.1$ and $\epsilon_l=0.2$. In our setting, the total wind turbine capacity accounts for 30\% of the total base load at the peak hour. Also, the covariance matrix $\Sigma_t$ is diagonal and the standard deviation for each element in $\boldsymbol{\omega}_t$ is equal to 30\% of the wind power forecast values. For the MDF bids, we assume that all the three load aggregators have the same bid for the VBs, i.e., $R_i^{min}=-0.1,R_i^{max}=0.1,E_i^{min}=-0.6,E_i^{max}=0.6,i=5,7,9$. However, the available service periods are different, which are [3:00,9:00], [10:00,16:00] and [17:00,23:00] for buses 5, 7 and 9, respectively. Moreover, the three load aggregators submit the same values of $\gamma^P$ and $\gamma^E$, while we vary them from 110 to 910 for $\gamma^P$ and from 100 to 2500 for $\gamma^E$. We show the absolute values for $\alpha_i^{E-}$ and $\alpha_i^{R+}$ with the MDF bids with different reward parameters in Fig. \ref{mdf_fig:333}.

From the figure, we see that at all the buses, the absolute cleared values for $\alpha^{E-}$ and $\alpha^{R+}$ are non-increasing as $\gamma^P$ and $\gamma^E$ increase. However, the clearing results start to decrease at different \textit{critical} values of the bidding parameters. It can be seen that the MDF bid at bus 7 is the most valuable one to the operator compared with the other two buses, because its clearing result is always no less than those of the other two. This result is observed because different MDF bids are located at different buses in the transmission network and more importantly, the service period of the load aggregator at bus 7 can lead to the most significant reduction of the generation cost since it covers the peak load period. Therefore, the evaluation of different MDF bids can be obtained by solving the market clearing problem.

\subsection{Beyond the Normal Wind Forecast Errors}
We also test the reliability of our models by simulating different distributions of wind turbines' output, i.e., Laplace, Logistic, Normal, Uniform and Weibull distributions \cite{xie2017distributionally}. We generate
$10^5$ samples from each distribution and show the probability of the line flow violation at the peak load hour in Table \ref{mdf_table:666}. From the table, we find that with the risk parameter $\epsilon_l=0.2$, the probabilities of the line flow violation for most distributions are very small (less than 6\%), while it is slightly higher in the case of Logistic distribution. However, the robustness of the solution can be improved by leveraging other techniques \cite{zhang2017distributionally,summers2014stochastic,roald2015security,xie2017distributionally} such as distributionally robust methods.  
\begin{table}[!t]
	\caption{Probability of line flow violation of samples drawn from different distributions }
	\label{mdf_table:666}
	\centering
	\begin{tabular}{cccc}
		\hline
		Distribution & Prob. of violation & Distribution & Prob. of violation\\
		\hline
		Laplace & 0.055  & Logistic & 0.318 \\
		\hline
		Normal & 0.039  & Uniform & 0.020 \\
		\hline
		Weibull & 0.038  &  &   \\
		\hline
	\end{tabular}
\end{table}



\section{Conclusions}\label{mdf:sec5}
In this paper, we propose a day-ahead market in the transmission network for the load aggregators with MDF. The aggregate load flexibilities are modeled by VBs. Each load aggregator is also required to submit the key parameters in the reward function and the market operator can accept a portion of each bid. The wind power uncertainty is captured by imposing the chance constraints and the chance constrained market clearing problem is equivalently reformulated into an SOCP. From the case studies, we show that the MDF bids can effectively help flatten the LMPs. Moreover, we show that the market clearing process can help the operator evaluate different load flexibilities by integrating the MDF bids with the ED problem.

\section*{Acknowledgement}
This work was supported by the Hong Kong
Research Grants Councils General Research Fund under Project 16210215.

%

\vspace{-0.0cm}

\ifCLASSOPTIONcaptionsoff
  \newpage
\fi

\end{document}